\documentclass[twocolumn,aps,prb,superscriptaddress]{revtex4-2}
\usepackage[hidelinks]{hyperref}
\usepackage{color}
\usepackage{amsmath,amssymb}
\usepackage{bm}
\usepackage{pifont}
\usepackage{amssymb}  
\usepackage{bbold}
\usepackage{float}
\usepackage[normalem]{ulem}
\usepackage[utf8]{inputenc}
\usepackage{textgreek}
\usepackage[labelfont=bf]{caption} 
\usepackage{tikz}
\usetikzlibrary{shapes.geometric, arrows.meta, positioning, shadows, backgrounds}
\usepackage{bbold}
\usepackage{makecell}
\usepackage{pifont}   
\usepackage{dcolumn}  
\usepackage{bm}       
\usepackage{multirow} 
\usepackage{placeins}
\usepackage{mathtools}
\usepackage{subfigure}
\usepackage{amsmath}  
\usepackage{dblfloatfix}
\usepackage{tikz}
\usetikzlibrary{quantikz}

\usepackage{adjustbox} 
\usepackage[normalem]{ulem}
\usepackage{graphicx} 

\usepackage{breqn}
\usepackage{mathrsfs}

\usepackage{multirow}
\usepackage[margin=1in]{geometry}
\usepackage{subcaption}
\usepackage{booktabs}

\begin{document}

\title{Beyond Trotterization: Variational Product Formulas for Quantum Simulation}
\author{Ibsal Assi}
\affiliation{Department of Physics and Physical Oceanography$,$
Memorial University of Newfoundland and Labrador$,$ St. John’s$,$ Newfoundland $\&$ Labrador$,$ Canada A1B 3X7}
\author{Michael Vogl}
\affiliation{Physics Department$,$ King Fahd University of Petroleum $\&$ Minerals$,$ Dhahran 31261$,$ Saudi Arabia}
\affiliation{Interdisciplinary Research Center (IRC) for Advanced Quantum Computing (AQC) $,$ KFUPM$,$ Dhahran$,$ Saudi Arabia}

\author{Meenu Kumari}
\affiliation{Digital Technologies$,$ National Research Council Canada}
\affiliation{Perimeter Institute for Theoretical Physics$,$ Waterloo ON N2L 2Y5$,$ Canada}
\affiliation{Institute for Quantum Computing$,$ University of Waterloo$,$ Ontario N2L 3G1$,$ Canada}
\author{J. P. F. LeBlanc}
\affiliation{Department of Physics and Physical Oceanography$,$
Memorial University of Newfoundland and Labrador$,$ St. John’s$,$ Newfoundland $\&$ Labrador$,$ Canada A1B 3X7}
\affiliation{Compute Everything Technologies Ltd.$,$ St. John's$,$ Newfoundland $\&$ Labrador$,$ Canada}
\date{\today}

\begin{abstract}
We propose a variational alternative to the Trotter-Suzuki decomposition that provides greater control over errors while preserving the unitary structure of time evolution. The variational parameters in our ansatz are derived from a global action principle, where Euler-Lagrange equations govern their optimal dynamics. Unlike conventional wavefunction-based variational methods, our approach specifically targets the time evolution operation and this allows a single set of optimized parameters to be applied to any initial state for a fixed Hamiltonian avoiding costly optimization procedures.  Our method outperforms the standard Trotter-Suzuki formulas, typically achieving higher accuracy than higher-order Suzuki schemes. This translates directly to quantum computing applications, where it enables the design of quantum circuits with fewer gates which reduces noise and improves precision. Although we focus on quantum dynamics, the method is broadly applicable to problems involving general time-evolution operators. Applied to various model Hamiltonians, our approach reduces errors by factors of 2 to 5 compared to Trotter-Suzuki decompositions, demonstrating its promise for accurate quantum simulation with improved efficiency. In certain cases, the variational ansatz achieves higher accuracy than more complex higher-order Suzuki formulas while reducing the gate count by nearly half within a single circuit layer. Furthermore, we derive approximate analytical expressions for the variational parameters up to cubic order in time, valid for generic Hamiltonians. These approximations enable long-time quantum simulations with improved accuracy over equivalent Suzuki decompositions, providing ready-to-use evolution formulas that match Suzuki's gate complexity while delivering better performance.
\end{abstract}

\maketitle

\section{Introduction}
Many-body quantum systems are described by Hilbert spaces whose dimensions grow exponentially with system size, such as the number of particles or spins. This rapid growth severely limits the ability to simulate realistic models on classical computers. For instance, exact diagonalization of one-, two-, or three-dimensional spin systems is constrained by the exponential increase in the number of sites and spin configurations \cite{ED1,ED2,ED3}. Quantum computers have been proposed as a promising solution to this challenge, offering the potential for significant computational advantages over classical approaches \cite{feynman1982,lloyd1996,Berry2007}, and have therefore attracted considerable interest \cite{QC_ex_1,QC_ex_2,QC_ex_3,QC_ex_4}. A key task in this context is the accurate simulation of quantum time evolution under a given time-independent Hamiltonian $H$, as this is essential for understanding the dynamical and static properties of quantum systems.

In digital quantum simulation (DQS), the time evolution operator of $H$, $U(t)=e^{-iHt}$, is implemented by decomposing it into a sequence of discrete unitary operations, or quantum gates. Typically, this operator cannot be expressed directly in terms of the elementary gates available on a quantum computer. The Trotter-Suzuki (TS) decomposition addresses this challenge by systematically approximating $U(t)$ as a product of simpler unitaries, each of which can be mapped into a sequence of gate operations in the corresponding quantum circuit. In the case of long time evolutions, one ideally chooses a valid trotterization for a small time step then repeat the same unitary multiple times until the full time interval is covered. This approximation introduces a controllable error that can be reduced by decreasing the time step, providing a flexible and systematic framework for digital quantum simulation \cite{trotter1959product,suzuki1991general}. 

However, the practical utility of TS methods on near-term quantum devices is severely limited by the large quantum circuit depths required for accurate long-time evolution. Indeed, accuracy depends on local error bounds per time step; obtaining highly accurate results requires taking many small time steps, which yields deep quantum circuits (QCs). Often, this results in projected computation times that exceed the coherence times of the current Noisy Intermediate-Scale Quantum (NISQ) processors \cite{NIQS_1,NIQS_2}. This problem necessitates the use of alternative (variational) methods that permit the design of shallower QCs. 

A significant body of work has explored variational principles applied directly to a given quantum state. Methods like the Variational Quantum Eigensolver (VQE) \cite{VQE1,VQE2,VQE3,VQE4}, adapted for dynamics, and the McLachlan's variational principle minimize the error between the time derivative of a parameterized wavefunction ansatz $|\psi(\boldsymbol{\theta}(t))\rangle$ and the exact Schrödinger evolution $ -iH |\psi(\boldsymbol{\theta}(t))\rangle$ \cite{Yuan2019theoryofvariational,PRXQuantum.4.030319}. These methods have shown promise for simulating specific types of evolution, particularly for states slowly varying in a constrained manifold.

Nonetheless, wavefunction-based variational methods possess inherent limitations. Their accuracy is often state-dependent, optimized for a particular initial state rather than an entire unitary $U(t)$ and the full Hilbert space. This issue renders them unsuitable for applications requiring knowledge of the full time-evolution operator, such as simulating the dynamics of arbitrary states. Moreover, it complicates the implementation in quantum gates, or in algorithmic error suppression techniques like probabilistic error cancellation, where an exact unitary structure is crucial. Furthermore, they can be susceptible to local minima and optimization challenges inherent in variational quantum algorithms \cite{Cerezo2021-nm,BarrenPlateaus}.

In this work, we overcome some of these limitations by introducing a global variational principle that applies directly to the unitary time-evolution operator itself~\cite{vogl2025variational}. Our approach optimizes a parameterized unitary  $U(\boldsymbol{c}(t))$ to best satisfy the operator Schrödinger equation over an entire time interval. This yields a set of differential equations for the variational parameters $\boldsymbol{c}(t)$, given physically motivated initial conditions. This shift in perspective---from local, state-based error to global, operator-based error---represents a fundamental departure from both Trotter-Suzuki and wavefunction variational methods, leading to substantial improvements in accuracy and efficiency, as we shall demonstrate.

This paper is structured as follows. We begin by introducing the theoretical framework of our method and the different forms of the parametrized time-evolution ansatz. We then demonstrate its application on three representative systems: a two-level toy model for pedagogical clarity, the quantum Ising model (QIM) as a paradigmatic case in quantum simulation, and the non-integrable XXZ model with next-nearest-neighbor interactions to highlight the robustness of our method. Subsequently, we illustrate the construction of the quantum circuit for our variational ansatz using the QIM as a concrete example. Next, we discuss approximate variational techniques that reduce computational costs, including the derivation of analytical expressions for the variational parameters which offer improved accuracy over Trotter--Suzuki formulas. We conclude with a summary of our results and an outlook on future research directions.

\section{Methodology}
In this section, we describe how to derive the equations of motion for the variational parameters of a general time-evolution ansatz using an action principle. We then introduce several parameterized product formulas for the time-evolution operator, where each unitary factor can be implemented as a sequence of elementary quantum gates. Finally, we discuss how long-time simulations can be carried out using stroboscopic time steps, making our approach directly applicable to quantum circuit implementations. 

\subsection{Quantum dynamics from action principles}
The exact time evolution operator $U(t)$ for a given Hamiltonian $H$, which we assume as time-independent throughout this work, fulfills the Schrodinger equation (SE),
\begin{equation}
\label{eqn:SchEq}
    i\partial_tU=HU.
\end{equation}
We have adopted the unit system where $\hbar=1$ throughout this work. Consequently, for the remainder of this work, all quantities are expressed in dimensionless units; equivalently, time is measured in units of inverse energy. One way to derive Eq. \eqref{eqn:SchEq} is via the action \cite{vogl2025variational} 
\begin{equation}
\label{eqn:action_S}
    S=\int dt \mathcal{L}_1=\int dt \, \mathrm{Tr} \left[ U^{\dagger} ( i \partial_{t} U - H U ) \right].
\end{equation}
The Schrödinger equation can be derived by minimizing the action with respect to the entries of $U^\dagger(t)$. As with any valid variational principle, this framework can be used to reduce computational complexity by restricting the dynamics to a carefully chosen ansatz, $U_{a}(t)$. In practice, $U_{a}(t)$ is parametrized using a small set of variational parameters---far fewer than the total number of independent entries in the full unitary matrix $U(t)$. Such a parametrization not only provides a computationally efficient representation of the time-evolution operator but can also offer additional physical insight into the system's dynamics. This approach is particularly relevant for quantum computing, where the goal is to implement $U$ on a quantum device using standard quantum gate operations. For example, in digital quantum simulations (DQS), one typically chooses an ansatz that both approximates the exact time-evolution operator and is naturally suited for implementation on a quantum circuit, as is the case with the well-known Trotter-Suzuki decomposition~\cite{trotter1959product,TS2}. Inspired by this structure, we adopt a similar strategy and express our ansatz for the time-evolution operator as a product of exponentials, as will be discussed in more detail in Section \ref{sec:Ansatz}. Crucially, our ansatz is parameterized by a set of time-dependent variational parameters, $\{c_j(t)\}$. By minimizing the action $S$ with respect to these parameters, we obtain the corresponding equations of motion (EOM) that govern their dynamics \cite{vogl2025variational}
\begin{equation}
\label{eqn:diff_eq_var_parms}
    \sum_{k} g_{jk} \dot{c}_k + i F_j = 0
\end{equation}
where $g_{jk} = \mathrm{Tr} \left[ \left( \frac{\partial U_{a}}{\partial c_j} \right)^\dagger \frac{\partial U_{a}}{\partial c_k} \right]$ 
is the quantum geometric tensor (QGT), which encodes the overlap between tangent vectors in the parameter space of the ansatz and determines the geometry of its evolution~\cite{Yuan2019theoryofvariational,QGT_2,vogl2025variational}, and 
$F_j = \mathrm{Tr} \left[ \left( \frac{\partial U_{a}}{\partial c_j} \right)^\dagger H U_a \right]$ 
is the generalized force term that represents the projection of the Hamiltonian's action onto the variational manifold, driving the time evolution of the parameters. The solutions to these equations can be obtained using standard numerical techniques, such as Runge-Kutta integration. In certain cases, it is even possible to derive exact or approximate analytical solutions for the variational parameters, as we will demonstrate in later sections.

It is important to note that the action introduced above is not unique. In fact, there exist infinitely many possible action principles that yield the exact dynamics and can serve as starting points for a given parameterized time-evolution ansatz. A few such alternatives are listed in Appendix~\ref{sec:alternative_S}. Throughout this work, we primarily focus on the action principle ($\mathcal{L}_1$) introduced above.
 
In the next section, we will discuss various forms of variational ansatz for the time-evolution operator that have direct applications to quantum computing.

\subsection{Time-evolution ansatz}
\label{sec:Ansatz}
In this section, we explore a special class of unitary ansatz that can be easily translated into a quantum circuit. As a first scenario, we assume that our Hamiltonian $H$ is decomposed into two non-commuting terms $A$ and $B$. Here, ideally $A$ or $B$ consist of a linear combination of terms that are mutually commuting. The goal is to approximate the time-evolution operator as a product of unitary operators made from $A$ and $B$ where each unitary can be implemented as a set of gate operations in a quantum circuit. For such an ansatz the only computational errors would be due to neglecting the nonzero commutator $[A,B]$. We consider the following variational ansatz for the time-evolution operator:
\begin{equation}
\label{eqn:ansatz_AB}
    U_a(t) = \prod_{j=0}^{\mathcal{M}} e^{i c_{2j}(t) A} e^{i c_{2j+1}(t) B},
\end{equation}
where $\mathcal{M}$ is an integer that controls the circuit depth and can be tuned to balance expressiveness and computational cost. The coefficients $\{c_k(t)\}$ are real-valued variational parameters that vanish at $t=0$, which ensures that $U_a(\tau)$ remains unitary as it is essential for quantum computing applications. 

Our next step is make a connection between the above general ansatz and Trotter-Suzuki decomposition. To do so, we start by doing Taylor expansion around $t=0$, we have
\begin{equation}
    \left.\frac{\partial U_a}{\partial c_{2j}}\right|_{\mathbf{c}=0}=iA, \quad \left.\frac{\partial U_a}{\partial c_{2j+1}}\right|_{\mathbf{c}=0}=iB
\end{equation}
Thus, the quantum geometric tensor becomes
\begin{equation}
    \left.g_{jk}\right|_{\mathbf{c}=0}=\begin{cases}
        \mathrm{Tr}[A^2] & \text{if $j$ and $k$ are even},\\
        \mathrm{Tr}[B^2] & \text{if $j$ and $k$ are odd},\\
        \mathrm{Tr}[AB] & \text{else}.
    \end{cases}
\end{equation}
and the generalized force at $t=0$ is
\begin{equation}
    \left. F_j\right|_{\mathbf{c}=0}=\begin{cases}
        -i\mathrm{Tr}\left[A(A+B)\right] & \text{if $j$ is even},\\
        -i\mathrm{Tr}\left[B(A+B)\right] & \text{if $j$ is odd}.
    \end{cases}
\end{equation}
Defining $S_e(t)=\sum_j c_{2j}(t)$ and $S_o(t)=\sum_j c_{2j+1}(t)$, and using Eq. \eqref{eqn:diff_eq_var_parms} at $t=0$, we obtain:
\begin{equation}
\mathrm{Tr}\left[A^2\right](\dot{S}_e(0)+1)+\mathrm{Tr}\left[AB\right](\dot{S}_o(0)+1)=0,
\end{equation}
\begin{equation}
\mathrm{Tr}\left[AB\right](\dot{S}_e(0)+1)+\mathrm{Tr}\left[B^2\right](\dot{S}_o(0)+1)=0,
\end{equation}
for even $j$ and odd $j$, respectively. The above system of algebraic equations have solutions $\dot{S}_e(0)=\dot{S}_o(0)=-1$. Thus, we have $\sum_j c_{2j}(t)=\sum_j c_{2j+1}(t)=-t+\mathcal{O}(t^2)$. Due to the symmetry of the ansatz, we have $c_{2j}(t)=c_{2j+1}(t)=-\frac{t}{\mathcal{M}+1}+\mathcal{O}(t^2)$, giving
\begin{equation}
    U_a(t)=\Big[e^{-i\frac{t}{\mathcal{M}+1}A} e^{-i\frac{t}{\mathcal{M}+1}B}\Big]^{\mathcal{M}+1}+\mathcal{O}\Big(\frac{t^2}{\mathcal{M}+1}\Big)
\end{equation}
which aligns well with the Trotter-Suzuki decomposition. Consequently, in the limit of $\mathcal{M}\to\infty$, or equivalently $t/(\mathcal{M}+1)\to 0$, $U_a(t)$ recovers the full (exact) time-evolution.

Having established the connection to TS decomposition, we move forward to further discuss the full expression in Eq. \eqref{eqn:ansatz_AB}. For finite-dimensional quantum systems such as spin models, there exists a fundamental mathematical guarantee that for sufficiently large but finite $\mathcal{M}$, this product (\eqref{eqn:ansatz_AB}) can exactly equal $U(t)$ for all times $t$. This follows from Lie-algebraic considerations: the operators $iA$ and $iB$ generate a finite-dimensional matrix algebra (a closed algebraic structure under commutation), and the time evolution operator $U(t)$ traces a continuous path within the corresponding matrix Lie group (the set of all possible unitary evolutions generated by this algebra). Since the generated Lie group forms a smooth, finite-dimensional manifold, and our product formula provides a parameterization with $2\mathcal{M}+1$ free parameters, once $\mathcal{M}$ is large enough that the number of parameters equals or exceeds the dimension of this manifold, exact representation becomes possible \cite{Hall2013,Serre2009}. For finite-dimensional spin systems, the required $\mathcal{M}$ is bounded by the dimension of the Lie algebra generated by $A$ and $B$, which is itself bounded by $\mathcal{D}^2-1$ where $\mathcal{D}$ is the Hilbert space dimension \cite{Kirillov2008}. This theoretical foundation ensures that variational optimization of the coefficients $c_k(t)$ can recover the exact quantum dynamics for finite $\mathcal{M}$.
 
One should also note that, because $A$ and $B$ consist of sums of mutually commuting operators, the exponentials in Eq.~\eqref{eqn:ansatz_AB} can be simplified further. Writing \(A=\sum_j \alpha_j P_j\) with \([P_j,P_k]=0\), one obtains
\begin{equation}
e^{i c_0 A}
= \prod_j e^{i c_0 \alpha_j P_j}.
\end{equation}
An identical simplification holds for exponentials involving $B$. For qubit Hamiltonians, $P_j$ are Pauli strings and by noting that $P_j^2=1$,  the above equation simplifies to product of matrices leading to a substantial reduction in computational cost. This factorization is useful for various analytical treatments as well as in certain numerical algorithms like the time evolving block decimation (TEBD) for matrix product states \cite{TEBD,MPS}.

Our variational ansatz is expected to outperform equivalent TS decompositions because it employs action principle that dynamically optimizes parameters ${c_k(t)}$ to satisfy the Schrödinger equation, whereas TS formulas follow fixed mathematical constructions. This physical optimization leads to enhanced accuracy for specific model Hamiltonians.

The simplest nontrivial instance of this ansatz consists of only two exponentials:
\begin{equation}
\label{eqn:ansatz_V1}
    U_a(t) = e^{i c_0(t) A} e^{i c_1(t) B},
\end{equation}
corresponding to a single layer of alternating unitaries generated by $A$ and $B$. We observe the similarity in structure to the first order Trotter-Suzuki (TS) formula $U_{\rm TS}^{(1)}(t)=e^{-it A} e^{-it B}$. Using Eq. \eqref{eqn:diff_eq_var_parms}, we find the equations of motion for the variational parameters $c_{0,1}(t)$ to be
\begin{equation}
    \mathrm{Tr}[A^2](1+\dot{c}_0) + \mathrm{Tr}[AB](1+\dot{c}_1)= 0
\end{equation}
\begin{equation}
    \mathrm{Tr}[AB](1+\dot{c}_0) +\mathrm{Tr}[B^2]\dot{c}_1 +\mathrm{Tr}\big[B\tilde{B}\big] = 0
\end{equation}
where $\tilde{B}=e^{-i c_0(t) A} B e^{i c_0(t) A}$.
Note that all time-independent traces appearing in the above equations can be evaluated analytically for arbitrary system sizes and Hamiltonian parameters, as demonstrated for the Ising and XXZ models in Appendix~\ref{app:traces}. However, the dynamical trace $\mathrm{Tr}\big[B\tilde{B}\big]$ is generally nontrivial to obtain analytically and is therefore computed numerically for more general models. Accordingly, we rely on numerical solutions of the above system of differential equations. For small $c_0(t)$, one may approximate $\tilde{B}\approx B$, in which case the system admits the analytic solutions $c_0(t)=c_1(t)=-t$, corresponding to the well-known Trotter–Suzuki parameters. This serves as a useful consistency check, since in the small-$t$ limit the variational parameters should recover the Trotter–Suzuki coefficients. 

It should be noted that the performance of the ansatz in Eq.~\eqref{eqn:ansatz_V1} depends on its operator ordering. For operators $A$ and $B$, the two natural choices are the AB ordering ($U_a = e^{i c_0 A} e^{i c_1 B}$) and the BA ordering ($U_a = e^{i c_0 B} e^{i c_1 A}$). While the variational principle optimizes the parameters for either choice, one ordering typically yields a more accurate solution. The preferred choice can be determined by comparing the residual $R=\|i\dot{U}_a-HU_a\|_F^2=\sqrt{\mathrm{Tr}\left[H^2\right]-\sum_jf_j\dot{\theta}_j}$, where $||X||_F=\sqrt{\mathrm{Tr}\left[XX^\dagger\right]}$ denotes the Frobenius norm of $X$, with the smaller value indicating better performance. For this simple ansatz one finds this residual for the AB ordering to be
\begin{align} 
\label{eqn:R_AB}
        R_{\rm AB}=&(\dot{c}_0+1)^2\mathrm{Tr}[A^2]+(\dot{c}_1^2+1)\mathrm{Tr}[B^2]\nonumber \\
        &2(\dot{c}_0+1)(\dot{c}_1+1)\mathrm{Tr}[AB]+2\dot{c}_1\mathrm{Tr}[B\tilde{B}]
\end{align}
One can easily find $R_{\rm BA}$ by the swap $A\leftrightarrow B$ and noting that the corresponding variational parameters change with operator ordering. We define the ordering parameter $\Delta={\rm sign}(R_{\rm AB}-R_{\rm BA})$ and if $\Delta=-1$, then the AB operator ordering is more accurate while for $\Delta=+1$, the BA is the correct operator form to proceed with. In Section \ref{sec:approx_analytic_vars_AB}, we will provide approximate analytic expression for $\Delta$ in terms of analytic traces involving $A$ and $B$. The procedure outlined here can be extended to more complicated ansatz and Hamiltonian structures if necessary.

Aside from the operator ordering, we recall the second order (symmetric) TS formula
\begin{equation}
\label{eqn:symmTS}
U_{\rm TS}^{(2)}(t)=e^{-\frac{it}{2}A}e^{-it B}e^{-\frac{it}{2}A}.
\end{equation}
which we rely on to choose our next variational ansatz of the form
\begin{equation}
\label{eqn:ansatz_V2}
U_a(t)=e^{ic_0(t)A}e^{ic_1(t)B}e^{ic_2(t)A}.
\end{equation}
where we took $c_{j>2}(t)=0$ in Eq. \eqref{eqn:ansatz_AB}. The variational parameters $\{c_0(t),c_1(t),c_2(t)\}$ are found by solving the differential equations in Eq. \eqref{eqn:diff_eq_var_parms} (or alternatively the explicit forms in Appendix \ref{app:EOM}), or the ones corresponding to the other action principles discussed in Appendix \ref{sec:alternative_S}. Naturally, one can consider higher-order generalizations of this ansatz by including additional alternating exponentials, as illustrated below:
\begin{equation}
\label{eqn:ansatz_V3}
    U_a(t) = e^{i c_0(t) A} \, e^{i c_1(t) B} \, e^{i c_2(t) A} \, e^{i c_3(t) B}.
\end{equation}
and so on. Increasing the number of factors allows the ansatz to capture more complex dynamics and can improve upon standard Trotterization schemes. In particular, higher-order constructions can enable accurate simulations over significantly larger time steps, as we will demonstrate later.  

At first glance, adding more factors per layer appears to increase the number of quantum gates and, consequently, the circuit depth. However, the relationship is more subtle. While each layer becomes more complex, the ability to take larger time steps often reduces the total number of layers required to simulate the full evolution. Therefore, an appropriate measure is the \emph{effective circuit depth}, which should be assessed in terms of the total gate count over the entire evolution, rather than per layer alone.

Having established this trade-off in the context of Hamiltonians that can be written as $H = A + B$, it is natural to ask how the method extends to systems with more intricate structures. In many physically relevant models, the Hamiltonian cannot be decomposed into only two non-commuting parts, each composed of mutually commuting operators. In such cases, it is convenient to generalize the decomposition to three or more terms. 

As a concrete example, consider the XXZ Hamiltonian with both first- and second-nearest-neighbor interactions. This Hamiltonian can be partitioned into three components, $A$, $B$, and $C$, where $A$ contains only the $XX$ interaction terms, $B$ contains only the $YY$ interaction terms, and $C$ contains the Ising ($ZZ$) terms. Each of these terms consists solely of mutually commuting operators on the $N$-qubit system. Consequently, the corresponding exponentials $e^{i a_j(t) A}$, $e^{i b_j(t) B}$, and $e^{i c_j(t) C}$ can be factorized exactly into parallel quantum gates, making the ansatz readily implementable on a quantum circuit.  

A natural variational ansatz for this case is
\begin{equation}
\label{eqn:ansatzABC}
    U_a(t) = \prod_{j=1}^{\mathcal{M}} e^{i a_j(t) A} \, e^{i b_j(t) B} \, e^{i c_j(t) C},
\end{equation}
where the time-dependent variational parameters $\{a_j(t), b_j(t), c_j(t)\}$ are determined according to Eq.~\eqref{eqn:diff_eq_var_parms} (or using other Lagrangians as discussed in Appendix \ref{sec:alternative_S}). Such an ansatz $U_a(t)$ can be easily mapped into a sequence of quantum gate operations for quantum computing applications.   

Alternatively, one can employ the ansatz in Eq.~\eqref{eqn:ansatz_AB} in two stages.
In the first stage, the Hamiltonian is partitioned into two blocks, $A$ and $B+C$, leading to the ansatz

\begin{equation}
\label{eqn:ansatzABC_2steps_1}
U_a(t) = \prod_{j=0}^{\mathcal{M}} e^{i c_{2j}(t) A}, e^{i c_{2j+1}(t) (B+C)}.
\end{equation}

In the second stage, each factor $e^{i c_{2j+1}(t) (B+C)}$ is further decomposed as an alternating product of exponentials of $B$ and $C$, according to Eq.~\eqref{eqn:ansatz_AB}:
\begin{equation}
\label{eqn:ansatzABC_2steps_2}
e^{i c_{2j+1}(t) (B+C)}
= \prod_{k=0}^{\mathcal{N}} e^{i d_{2k}^{(j)}(t) B}, e^{i d_{2k+1}^{(j)}(t) C}.
\end{equation}

Care must be taken at this stage, since $(B+C)$ now plays the role of an effective Hamiltonian and $-c_{2j+1}(t)$ serves as the evolution time. This distinction is essential when deriving the corresponding equations of motion. Combining the two steps, the full ansatz takes the form
\begin{equation}
\label{eqn:ansatzABC_2steps_3}
U_a(t)
= \prod_{j=0}^{\mathcal{M}}
\prod_{k=0}^{\mathcal{N}}
e^{i c_{2j}(t) A}
e^{i d_{2k}^{(j)}(t) B}
e^{i d_{2k+1}^{(j)}(t) C}.
\end{equation}
which is directly implementable on a quantum computer. 

We will illustrate the performance of both the splitting strategies using the XXZ model in Sec.~\ref{sec:applications}. We will show that both the methods outperform the standard Trotter–Suzuki decomposition and yield nearly identical accuracy at short evolution times $t$. At longer times, however, the single-step splitting achieves lower errors. This improvement stems from its greater flexibility, as it optimizes a larger set of variational parameters simultaneously, whereas the two-step procedure effectively constrains the number of free parameters at each stage. Further numerical results and supporting analysis are presented in Sec.~\ref{sec:applications}.

This construction can be extended straightforwardly to Hamiltonians that decompose into more than three mutually commuting blocks. The same procedure can be readily applied to such cases. 

\subsection{General Applicability and Scalability of the Method}
\label{sec:generalapplicability}
While the variational principle is completely general, its practical implementation requires the evaluation of the quantum geometric tensor $g_{jk}$ and the generalized force $F_{j}$, defined in terms of traces over a $2^N$ dimensional Hilbert space. At first glance, these quantities seem to involve operators that act on the full space, suggesting an exponential computational cost. However, their structure often permits efficient evaluations using two complementary strategies: (i) classical algorithms that exploit operator sparsity, and (ii) hybrid quantum-classical algorithms that leverage the capabilities of near term quantum devices. Later in this paper, we demonstrate an extreme case, the quantum Ising model, where the traces can be evaluated exactly in a closed form regardless of the system size and coupling parameters as illustrated for the ansatzes Eqs. \eqref{eqn:ansatz_V2} and \eqref{eqn:ansatz_V3} in Appendix \ref{app:traces}. However, for general Hamiltonians such analytic calculations could be difficult to obtain and may not be feasible for more complex ansatzes.  In these cases the methods outlined in this section become essential. 

Various classical algorithms can be used to evaluate the traces even for large number of qubits. The exponentials appearing in the time-evolution ansatz are products of exponentials of local Hamiltonians. Because the terms within each exponential are sum of mutually commuting local operators, each exponential can be expressed as a matrix product operator (MPO) with small bond dimension $\chi$ that is independent of the system size $N$ \cite{MPS}. The operators inside $g_{jk}$ and $F_j$ are products of such MPOs, and the product of two MPOs is again an MPO with bond dimension at most the product of the individual bond dimensions. The trace of an MPO is obtained by contracting the physical indices, which reduces to multiplying a chain of $\chi\times\chi$ transfer matrices, an operation that scales as $\mathcal{O}(N\chi^3)$ and thus is linear in $N$ \cite{MPS}. This approach is applicable to a wide family of model Hamiltonians common in condensed matter physics \cite{XXZ_1,Pfeuty1970,Kopec1989}.

In the scenario where the bond dimensions are large where classical computations become prohibitive, one needs to switch to quantum computers. The traces in $g_{jk}$ and $F_j$ can be expressed as sums of traces of unitary operators. For instance, let's consider the quantum metric component $g_{02}=\mathrm{Tr}\left[Ae^{ic_1B}Ae^{-ic_1B}\right]$ appearing in the quantum Ising model where $A = \frac{h_x}{2} \sum_{j=1}^N \sigma_j^x$ and $B = \frac{J}{4}\sum_{j=1}^{N-1} \sigma_j^z \sigma_{j+1}^z + \frac{h_z}{2}\sum_{j=1}^N \sigma_j^z$. One can simply write
\begin{equation}
    g_{02}=\frac{h_x^2}{4}\sum_{j,k=1}^N\mathrm{Tr}\left[\sigma_j^xe^{ic_1B}\sigma_k^xe^{-ic_1B}\right]
\end{equation}
where $\sigma_j^xe^{ic_1B}\sigma_k^xe^{-ic_1B}$ is unitary and can be expressed as a quantum circuit. Note that due to commutation relations, this unitary can be simplified because all of the qubits that are not related to qubits $j$ and $k$ will have cancelling contributions due to the opposite signs in the exponentials. To find the trace of each term, one performs Hadamard tests with a system initialized in a sequence of basis states (or random states), yielding a trace up to $2^N$ factor. Summing all the $N^2$ terms reconstructs $g_{02}$. In principle, this applies to general $g$ and $F$: they are linear combinations of traces of unitaries each measurable via Hadamard test \cite{Hadamard_1,Hadamard_2}. This approach is rooted in the linear combinations of unitaries (LCU) framework \cite{LCU}. Together, the classical and quantum strategies provide a comprehensive toolkit for scaling the variational method to large systems, ensuring its applicability across a broad range of Hamiltonians. 

\subsection{Repeated evolution over multiple time periods}
In digital quantum simulation (DQS), it is often necessary to simulate systems over evolution times $t$ that exceed the time scales where a given ansatz remains accurate. In other words, $t$ is large enough that the normalized error $\frac{1}{2\sqrt{\mathcal{D}}} | e^{-itH} - U_a(t) |_F$ exceeds the acceptable error bound, where $\mathcal{D}$ is the Hilbert‑space dimension. Within the standard Trotter-Suzuki (TS) framework, the time-evolution operator in such a scenario can be approximated as
\begin{equation}
    e^{-i t (A + B)} \approx \left[e^{-i \tau A} e^{-i \tau B}\right]^n,
\end{equation}
where $\tau$ is a small time step and $n = \left\lfloor \frac{t}{\tau} \right\rfloor$. This decomposition becomes exact in the limit $\tau \to 0$ implying $n \to \infty$. For finite $n$, the error is bounded by $\frac{n}{2\sqrt{\mathcal{D}}}||e^{-i\tau H}-e^{-i \tau A} e^{-i \tau B}||_F$.  

In our approach, a similar strategy can be adopted by obtaining the variational parameters at $t=\tau$ by simply integrating the equations of motion Eq.~\eqref{eqn:diff_eq_var_parms} for $t\in[0,\tau]$. For instance, using the two-parameter ansatz of Eq.~\eqref{eqn:ansatz_V1}, the evolution over a total time $t$ can be approximated as
\begin{equation}
\label{eqn:U_strop_simple}
    e^{-i t (A + B)} \approx \left[e^{-i c_0(\tau) A} e^{-i c_1(\tau) B}\right]^n.
\end{equation}
More generally, for a higher-order ansatz $U_a(t)$, we simply replace the product above with the full variational form,
\begin{equation}
\label{eqn:U_strop_general}
    e^{-i t (A + B)} \approx \left[U_a(t=\tau)\right]^n,
\end{equation}
where $\tau=t/n$. This procedure enables accurate simulation of state evolution or observables over long times using a fixed, optimized ansatz at a single time step $\tau$. We will show in Sec.~\ref{sec:applications} that this approach can achieve higher accuracy when compared to the conventional TS methods. Equivalently, for the same total gate count, our scheme requires fewer layers in the quantum circuit, leading to a more resource-efficient simulation.

As discussed above, there is considerable flexibility in the choice of ansatz and the ordering of operators. This flexibility allows one to select the form that yields the smallest relative error for a given problem. While it is not feasible to explore all possible configurations here, our method can be readily applied to any ansatz that may be advantageous for specific models of interest.  

We summarize our algorithm in Fig.~\ref{fig:algorithm_schematics}. In the first step, we choose a particular time evolution ansatz $\boldsymbol{U}_a(\boldsymbol{c}(t))$ of interest that is implementable on quantum hardware. In the second step, the variational parameters $\boldsymbol{c}(t)$ are obtained by solving the equations of motion, Eq. \eqref{eqn:diff_eq_var_parms}. The output parameters $\boldsymbol{c}(t=\tau)$ are input for quantum simulations running at stroboscopic time steps of size $\tau$. This structure bears similarity to other algorithms based on McLachlan’s variational principle~\cite{McLachlanVarPrin}. However, our method differs in an essential way: the optimization is performed globally over the entire evolution operator rather than with respect to a specific quantum state in the Hilbert space.     

The implementation of our algorithm on actual quantum hardware is beyond the scope of the present work and is left for future investigation.

\begin{figure}[t]
\centering
\resizebox{\columnwidth}{!}{%
\begin{tikzpicture}[
    node distance=1.5cm and 1.2cm,
    every node/.style={font=\small},
    >=Stealth,
    init/.style={
        ellipse,
        draw,
        fill=blue!10,
        minimum height=0.9cm,
        minimum width=2.2cm,
        align=center,
        drop shadow
    },
    gf/.style={
        diamond,
        draw,
        fill=yellow!15,
        aspect=2,
        minimum height=1.2cm,
        minimum width=3.2cm,
        align=center,
        drop shadow
    },
    box/.style={
        draw,
        rounded corners=6pt,
        minimum height=1.2cm,
        minimum width=3.2cm,
        align=center,
        line width=0.6pt,
        fill=green!10,
        drop shadow
    },
    output/.style={
        draw,
        rounded corners=6pt,
        fill=orange!10,
        minimum height=1.2cm,
        minimum width=3.2cm,
        align=center,
        drop shadow
    }
]

\node[init] (ansatz) {Ansatz $U_a(c(t))$};

\node[gf, right=of ansatz] (gf) {Construct $g$ \& $F$};

\node[box, below=2.0cm of gf] (eom) {Integrate EOM\\ (Eq.~3); $c(0)=0$};

\node[output, left=of eom] (output) {Parameters $c(t)$};

\draw[->] (ansatz) -- (gf);
\draw[->] (gf) -- (eom);
\draw[->] (eom) -- (output);

\node[below=0.5cm of output, 
      draw=gray!40, 
      rounded corners=3pt, 
      fill=gray!5, 
      inner sep=4pt, 
      font=\footnotesize\itshape,
      align=center] (annotation) 
      {Used for quantum simulations};

\draw[->, gray!50, line width=0.4pt] 
    (annotation.north) -- ++(0,0.15) -| (output.south);

\begin{scope}[on background layer]
\fill[gray!5, rounded corners] 
    ([shift={(-0.4,-0.4)}]current bounding box.south west) 
    rectangle 
    ([shift={(0.4,0.4)}]current bounding box.north east);
\end{scope}

\end{tikzpicture}%
}
\caption{Flowchart of the variational quantum dynamics algorithm. The ansatz parameters $c(\tau)$ are evolved by integrating the equations of motion (Eq.~3).}
\label{fig:algorithm_schematics}
\end{figure}

\section{Applications to various models}
\label{sec:applications}
In this section, we demonstrate our method through three illustrative examples.
We begin with a simple two-level toy model, which offers analytical insight into the core ideas of our approach.
We then consider two many-body spin Hamiltonians that are generally non-integrable, except at special parameter values. For these systems, we analyze the behavior of the variational parameters and examine how the method performs as the system size increases toward the thermodynamic limit.

To quantify the accuracy of the approximate time-evolution ansatz $U_a(t)$ compared to the exact evolution $U(t)$, we define the following global error metric,
\begin{equation}
\label{eqn:error_metric}
\mathcal{E}_F(t)=\frac{1}{2\sqrt{\mathcal{D}}} \big\lVert U(t) - U_a(t) \big\rVert_F.
\end{equation}
The normalization factor ensures that $0\leq \mathcal{E}_F(t)\leq 1$. For stroboscopic dynamics with step size $\tau$ and an integer number of steps $n$, we use the analogous metric
\begin{equation}
\label{eqn:error_metric_strob}
\mathcal{E}_F^{\mathrm{Strob}}(\tau,n)=\frac{1}{2\sqrt{\mathcal{D}}} \big\lVert U(n\tau) - [U_a(\tau)]^n \big\rVert_F.
\end{equation}

\subsection{A simple two level system}
\begin{figure}
    \centering
    \includegraphics[width=\linewidth]{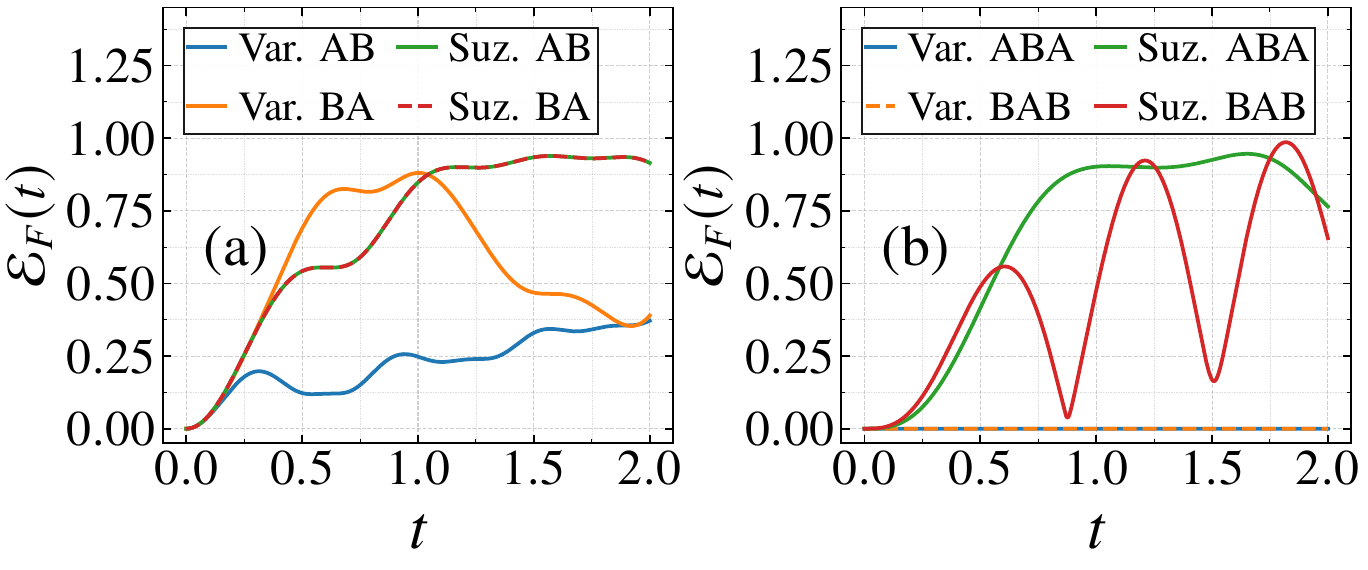}
    \caption{(a) A plot of the normalized Frobenius norm of the difference between the exact and the approximate time-evolution, $\mathcal{E}_F$ (Eq. \eqref{eqn:error_metric}), for the two-level model Eq. \eqref{eqn:2levelsH} using the variational ansatz $U_a(t)=e^{ic_0(t)A}e^{ic_1(t)B}$ in comparison with the first order Trotter-Suzuki (TS) decomposition for both operator orderings AB and BA. (b) Similar to (a), we employ the 3-exponential variational ansatz $U_a(t)=e^{ic_0(t)A}e^{ic_1(t)B}e^{ic_2(t)A}$ and compute the normalized Frobenius error against the symmetric TS formula Eq. \eqref{eqn:symmTS} for the operator orderings ABA and BAB. Parameters: $h_x=5$, and $h_z=2$.}
    \label{fig:Two_Levels_Frob_Norm}
\end{figure}

As a first example, we consider one of the simplest possible single particle Hamiltonians given by

\begin{equation}
\label{eqn:2levelsH}
    H=h_x\sigma_x+h_z\sigma_z=\frac{1}{2}\vec{B}\cdot\vec{\sigma}
\end{equation}
where $\vec{B}=(2h_x,0,2h_z)^T$ is a uniform magnetic field and $\vec{\sigma}=(\sigma_x,\sigma_y,\sigma_z)^T$. One can easily verify that the exact time evolution operator for this Hamiltonian is
\begin{equation}
    U_{\rm exact}(t)=\cos(\Omega t)-\frac{i}{\Omega}\sin(\Omega t)H
\end{equation}
where $\Omega=\sqrt{h_x^2+h_z^2}$. Now, if we let $A=h_x\sigma_x$ and $B=h_z\sigma_z$, then we use the ansatz in Eqs.\eqref{eqn:ansatz_V1}, the comparison against the equivalent first order Trotter-Suzuki (TS) formula is given in Fig. \ref{fig:Two_Levels_Frob_Norm}(a) at the different operator orderings where we simply $A \leftrightarrow B$. Meanwhile the TS results are independent of the operator orderings, we find that the variational results change with operator ordering in this case and that the case of $AB$ ordering our variational ansatz performed best globally compared to the TS results and the variational $BA$ ordering case. We repeated our calculations using the three-exponential ansatz in Eq.~\eqref{eqn:ansatz_V2}. As shown in Fig.~\ref{fig:Two_Levels_Frob_Norm}(b), the variational approach exhibits superior accuracy compared to the symmetric Trotter–Suzuki (TS) formula of Eq.~\eqref{eqn:symmTS} with error of order $10^{-7}$ that originates purely from numerical inaccuracies of the Runge–Kutta integrator (RK45 in Python).  

In fact, the variational approach exactly reproduces the true evolution in this case. To make this explicit, we simplify the ansatz in Eq.~\eqref{eqn:ansatz_V2} for the present system as follows:
\begin{equation}
    U_{\rm a}(t)=\alpha(t) I+i\beta(t) \sigma_x+i\gamma(t) \sigma_y+i\lambda(t)\sigma_z
\end{equation}
where $\alpha(t)=\cos(c_1h_z)\cos(h_x(c_0+c_2))$, $\beta(t)=\cos(c_1h_z)\sin(h_x(c_0+c_2))$, $\gamma(t)=\sin(c_1h_z)\sin(h_x(c_0-c_2))$, and $\lambda(t)=\sin(c_1h_z)\cos(h_x(c_0-c_2))$. By setting $U_{\rm a}(t)=U_{\rm exact}(t)$, we obtain
\begin{equation}
\label{eqn:c0_two_levels}
    c_0(t)=c_2(t)= \frac{1}{2 h_x} \tan^{-1} \left( -\frac{h_x \tan(\Omega t)}{\Omega} \right)
\end{equation}
\begin{equation}
\label{eqn:c1_two_levels}
    c_1(t) = \frac{1}{h_z} \sin^{-1} \left( -\frac{h_z \sin(\Omega t)}{\Omega} \right)
\end{equation}
as our unique solution. By doing Taylor expansion to linear order, we obtain $c_0(t)=c_2(t)\approx -\frac{t}{2}$ and $c_1(t)\approx-t$, these are the coefficients appearing the symmetric TS formula Eq. \eqref{eqn:symmTS} as one would expect in the limit $t\to0$. The above solutions cannot be obtained if one uses the simpler ansatz Eq. \eqref{eqn:ansatz_V1} for the fact that the equation $U_{\rm ansatz}(t)=U_{\rm exact}(t)$ has four equations with 2 unknowns in that case hard to satisfy all equations at once. For this system one would generally expect the need of 4 parameters in order to solve the above constraint consistently but luckily with three parameters we were able to find an analytic solution. The reasoning behind this lies in the field of Lie-Algebra following the arguments presented in the previous section. Using our variational approach, the equations of motion for our parameters are
\begin{equation}
\label{eqn:EOM_1Qubit_1}
g\dot{C}+b=0
\end{equation}
where
\begin{equation} 
\label{eqn:EOM_1Qubit_2}
g=\begin{bmatrix}
 h_x^{2} & 0 & h_x^{2} \cos(2 c_1 h_z) \\
0 &  h_z^{2} & 0 \\
 h_x^{2} \cos(2 c_1 h_z) & 0 &  h_x^{2}
\end{bmatrix}
\end{equation}
and
\begin{equation}
\label{eqn:EOM_1Qubit_3}
b=\begin{bmatrix}
h_x^{2} \\
h_z^{2} \cos(2 c_0 h_x) \\
h_x \left( h_x \cos(2 c_1 h_z) + h_z \sin(2 c_0 h_x) \sin(2 c_1 h_z) \right)
\end{bmatrix}
\end{equation}
where one can easily verify that the analytic solutions in Eqs. \eqref{eqn:c0_two_levels} and \eqref{eqn:c1_two_levels} do satisfy the equations of motion Eqs. \eqref{eqn:EOM_1Qubit_1}-\eqref{eqn:EOM_1Qubit_3}, indicating that the variational approach in this case recovers the analytic solutions which explains the high accuracy we obtained numerically shown in Fig. \ref{fig:Two_Levels_Frob_Norm}(b). Indeed, the alternative variational principles discussed in Appendix~\ref{sec:alternative_S} yield exactly the same result. This is expected, since all these principles are designed to reproduce the exact dynamics of the full unitary operator. In the present case of the three-exponential ansatz, the structure of the ansatz coincides with that of the exact unitary, which automatically ensures that all variational formulations produce identical parameters. Clearly, the variational approach provides more accuracy compared to using equivalent trotter-Suzuki formula. This will be clear later when using more complex models.   

For many-body Hamiltonians, such analytic solutions are generally difficult to obtain, as we will see in the following sections. Nevertheless, the overall trend remains the same: increasing the number of variational parameters typically improves the accuracy of the approximation. For instance, if the Hilbert space has dimension $\mathcal{D}$, one would, in principle, require at least $\mathcal{D}^2 - 1$ exponentials in the alternating ansatz of Eq.~\eqref{eqn:ansatz_AB} to fully capture the exact time evolution.  

The way these parameters are assigned strongly affects the resulting quantum circuit. Increasing the number of exponentials in the ansatz can significantly enhance expressivity, but it also introduces greater gate complexity. This trade-off can, however, lead to a shallower overall circuit, since fewer Trotter steps may be needed to achieve the same simulation time. Hence, the total gate complexity considering the full circuit will be smaller. In practice, the number of exponentials in the ansatz should be chosen to balance expressivity, hardware constraints, and the dominant sources of error in the quantum device.

\subsection{The Quantum Ising Model}
Our next example is the quantum Ising model (QIM), a standard benchmark for studying quantum dynamics and digital simulation techniques\cite{Lanyon2011,Peng2005,Barends2016,Lubasch2023,Lubasch2024}. It is one of the simplest many-body systems that captures essential quantum effects, combining local spin interactions with non-commuting transverse and longitudinal fields. Its structure naturally matches the capabilities of modern quantum hardware, where two-qubit $ZZ$ interactions and single-qubit rotations can be directly implemented. Moreover, in the absence of a longitudinal field ($h_z=0$), the model reduces to the integrable Transverse Field Ising Model (TFIM), which is analytically tractable \cite{Pfeuty1970}. This provides exact results for validation, while the full model with $h_z\neq0$ still exhibits rich phenomena such as quantum phase transitions and non-trivial entanglement dynamics \cite{Sachdev1999,Kopec1989}. These features make it an ideal testbed for exploring Trotterization and quantifying simulation errors. The Hamiltonian takes the form
\begin{equation}
\label{eqn:QIM}
H=J\sum_{j=1}S_j^zS_{j+1}^z+h_x\sum_{j=1}^NS_j^x+h_z\sum_{j=1}^NS_j^z
\end{equation}
where $\vec{S}_j=\frac{1}{2}\vec{\sigma}_j$ (i.e. we are using spin $1/2$ case). In this model, $J$ controls the strength of the nearest-neighbor Ising interactions, $h_x$ sets the magnitude of the transverse field that induces spin flips, and $h_z$ represents a longitudinal field that breaks integrability and adds additional complexity to the dynamics. Our goal is to split the Hamiltonian into different non-commuting terms with each term consisting of mutually commuting operators. One minimal choice is by writing $H=A+B$, where $A = h_x \sum_{j=1}^N S_j^x$ and $B = J \sum_{j=1}^{N-1} S_j^z S_{j+1}^z + h_z \sum_{j=1}^N S_j^z$. This decomposition allows the time-evolution ansatz to be directly implemented on a quantum computer since the individual exponentials of the form $e^{ic_{2j}(t) A}$ and $e^{ic_{2j+1}(t) B}$ are easily mapped into elementary gate operations. 

\begin{figure}
    \centering
    \includegraphics[width=\linewidth]{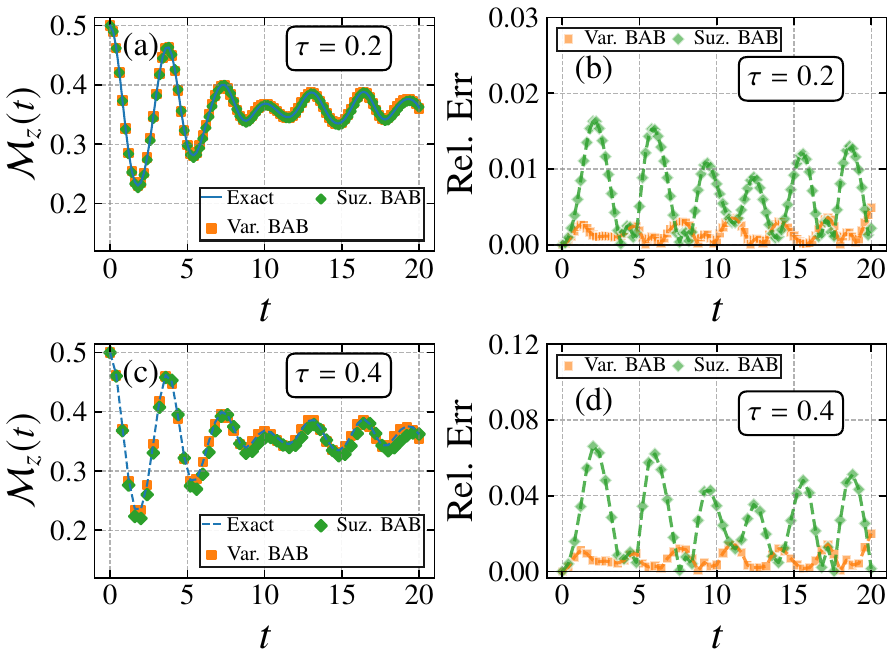}
    \caption{(a,c) Dynamical magnetization for the QIM [Eq. \eqref{eqn:QIM}] comparing exact calculations, the symmetric TS formula [Eq. \eqref{eqn:symmTS}], and our variational ansatz [Eq. \eqref{eqn:ansatz_V2}]. 
    (b,d) Relative error in magnetization comparing TS and variational methods with operator ordering BAB for stroboscopic time steps of sizes $\tau=0.2$ and $\tau=0.4$, respectively. 
    Parameters: $h_x = h_z = J = 1.0$, $N=10$, initial state $|\psi_0\rangle = |{\uparrow}\rangle^{\otimes N}$. Here $t$ is in units of $1/J$.}
\label{fig:MagAnalysis_Ising}
\end{figure}

Figures~\ref{fig:MagAnalysis_Ising}(a and c) show the dynamical magnetization, 
$\mathcal{M}_z(t) = \frac{1}{N} \sum_{j} \langle \psi(t) | S_j^z | \psi(t) \rangle$, 
for a system initialized with all spins polarized along the $z$-axis, using $h_x = h_z = 1.0$ and $N = 10$ qubits. 
The results are compared between exact calculations, the symmetric second-order Trotter-Suzuki (TS) formula, and our three-exponential variational ansatz in Eq.~\eqref{eqn:ansatz_V2}. 
In Figs.~\ref{fig:MagAnalysis_Ising}(b) and \ref{fig:MagAnalysis_Ising}(d), we plot the relative error in $\mathcal{M}_z(t)$ (defined as $|(\mathcal{M}_z^{\rm exact}(t)-\mathcal{M}_z^{\rm method}(t))/\mathcal{M}_z^{\rm exact}(t)|$) obtained via the TS and the variational ansatz for both stroboscopic time steps $\tau=0.2$ and $\tau=0.4$, respectively. We observe that the variational approach provides a better accuracy over the TS method with a factor of 4 or 5 times better, and this advantage becomes more pronounced at larger time steps $\tau$. While we discussed our results for the $\mathrm{BAB}$ operator ordering, the similar behavior was found for the $\mathrm{ABA}$ ordering except that the $\mathrm{BAB}$ ordering has the best accuracy for this model and the parameters chosen. These findings highlight the advantage of the variational approach, particularly in regimes where coarse time discretization would otherwise introduce substantial Trotter errors.

\begin{figure}
\centering
    \includegraphics[width=0.99\linewidth]{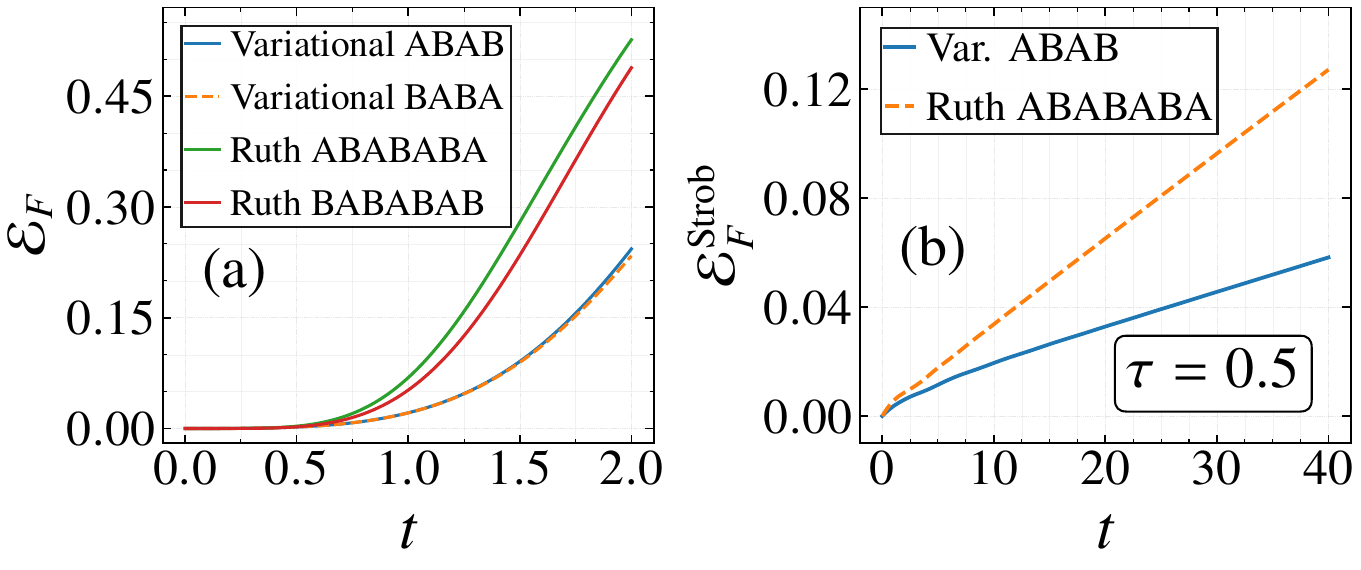}
\caption{(a) Normalized Frobenius norm difference between exact time evolution and approximate methods for the QIM [Eq. \eqref{eqn:QIM}]: variational ansatz [Eq. \eqref{eqn:ansatz_V3}] and Ruth's formula [Eq. \eqref{eqn:ruth}]. Operator orderings: variational (ABAB, BABA) and Ruth (ABABABA, BABABAB). (b) The stroboscopic error when using the variational ansatz and the Ruth decomposition with time step $\tau=0.5$. Parameters: $h_x=h_z=J=1.0$, and $N=10$. Here $t$ is in units of $1/J$.}
\label{fig:Ising_V4_frob}
\end{figure}

For further illustration, we compare our variational ansatz in Eq.~\eqref{eqn:ansatz_V3} with Ruth's fourth-order formula \cite{ruth1983canonical,forest1990fourth}, given by
\begin{equation}
\label{eqn:ruth}
    U_{\mathrm{Ruth}}(t) = U_{\mathrm{TS}}^{(2)}(pt)\, U_{\mathrm{TS}}^{(2)}(qt)\, U_{\mathrm{TS}}^{(2)}(pt),
\end{equation}
where $U_{\mathrm{TS}}^{(2)}(t)$ is the symmetric second-order Suzuki-Trotter formula in Eq.~\eqref{eqn:symmTS}, with $p = \frac{1}{2 - 2^{1/3}}$ and $q = 1 - 2p$. Figure~\ref{fig:Ising_V4_frob}(a) shows the normalized Frobenius norm results. The variational ansatz achieved lower error than the Ruth ansatz over the entire time interval $[0,2]$. At $t=1.0$, for example, the variational ansatz yielded approximately 1.5\% error, while the Ruth ansatz produced more than twice this error (over 3\%). These accuracies are obtained despite the fact that we are using fewer exponentials in our variational time-evolution formula. Specifically, Eq.~\eqref{eqn:ruth} requires seven exponentials per time step, while our variational ansatz uses only four.

To further demonstrate the practical advantage of our approach in quantum circuit implementations, we compare both ansatzes for long‑time simulations. As shown in Figure~\ref{fig:Ising_V4_frob}(b) (with time step $\tau=0.5$ and total time $t=40$), the stroboscopic error of the variational ansatz is consistently smaller than that of the Ruth ansatz. More importantly, the Ruth ansatz requires more exponentials per layer, leading to substantially deeper circuits. Table~\ref{tab:gate_counts} lists the gate counts per layer for each ansatz in terms of $R_x$, $R_z$, and CNOT gates, assuming $N$ qubits. For our simulation with $80$ layers (since $t/\tau = 80$), the total gate counts are obtained by multiplying the per‑layer numbers by $80$. For example, for $N=10$ qubits, the variational ansatz reduces $R_x$ gates by $50\%$, and $R_z$ and CNOT gates by about $33\%$ compared to the Ruth ansatz. Thus, our variational ansatz achieves superior accuracy and gate efficiency. We note that extending the variational framework to include more exponentials could further improve accuracy and enable larger time steps, leading to even shallower circuits~\cite{lee2025}.

\begin{table}[htbp]
\centering
\begin{tabular}{lcccccc}
\toprule
Ansatz & $\#A$ exp & $\#B$ exp & $R_x$ & $R_z$ & CNOT \\
\midrule
Variational & 2 & 2 & $2N$ & $4N-2$ & $4N-4$ \\
Ruth & 4 & 3 & $4N$ & $6N-3$ & $6N-6$ \\
\bottomrule
\end{tabular}
\caption{Comparison between the total number of $R_x$, $R_z$, and CNOT gates per layer appearing when implementing the variational ansatz in Eq. \eqref{eqn:ansatz_V3} for the $N$ qubits QIM versus the gate count when using the Ruth ansatz Eq. \eqref{eqn:ruth}.}
\label{tab:gate_counts}
\end{table}

\begin{figure}
    \centering
    \includegraphics[width=\linewidth]{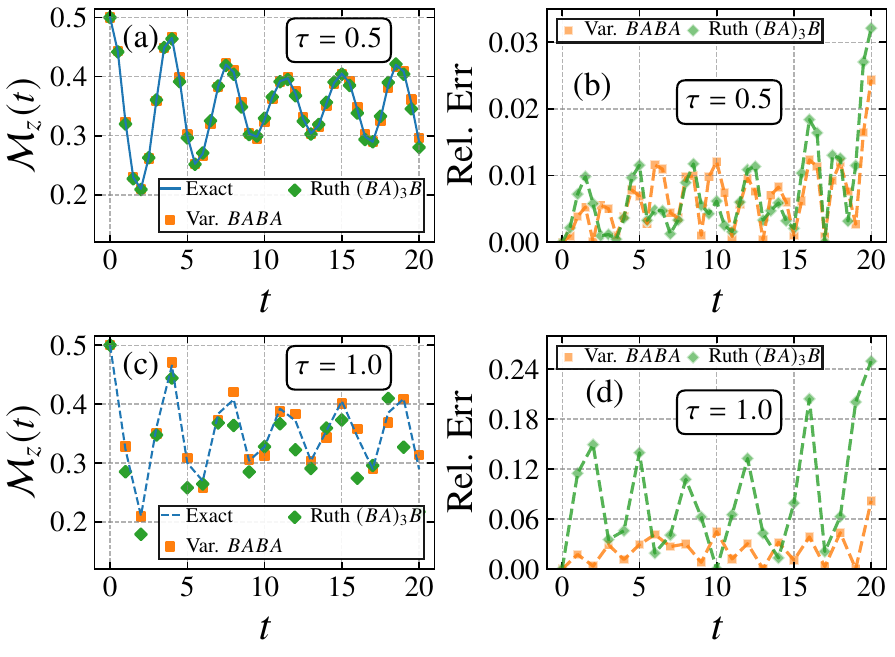}
    \caption{(a,c) The dynamical magnetization for the QIM [Eq. \eqref{eqn:QIM}] obtained via exact time evolution, Ruth formula [Eq. \eqref{eqn:ruth}], and four exponentials variational ansatz [Eq. \eqref{eqn:ansatz_V3}] for different operator orderings. For the variational ansatz with ordering BABA; for Ruth with operator ordering $(BA)_3B$ (denoting BABABAB). (b,d) The corresponding relative error in the dynamical magnetization at $\tau=0.5$ and $\tau=1.0$, respectively. Parameters: $h_x=h_z=J=1.0$, and $N=5$, with the initial state $|\psi_0\rangle=|\uparrow\rangle^{\otimes N}$. Here $t$ is in units of $1/J$.}
\label{fig:MagAnalysis_Ising_var_Ruth}
\end{figure}

In addition, We calculated the dynamical magnetization with the corresponding relative error at different time steps using the ansatz in Eq.~\eqref{eqn:ansatz_V3} and the Ruth formula Eq. ~\eqref{eqn:ruth} with BABA and BABABAB operator orderings, respectively, as shown in Fig. \ref{fig:MagAnalysis_Ising_var_Ruth}(a,c). Clearly, we obtain better accuracy using the variational ansatz compared to the Ruth formula and this sustains at larger time steps ($\tau$) as illustrated in the relative error in magnetization Fig. \ref{fig:MagAnalysis_Ising_var_Ruth}(b,d). This improvement is found despite the fact that we are using less exponentials in our parametrized ansatz. This further illustrates the utility of our approach for practical applications.

\begin{figure}
    \centering
    \includegraphics[width=\linewidth]{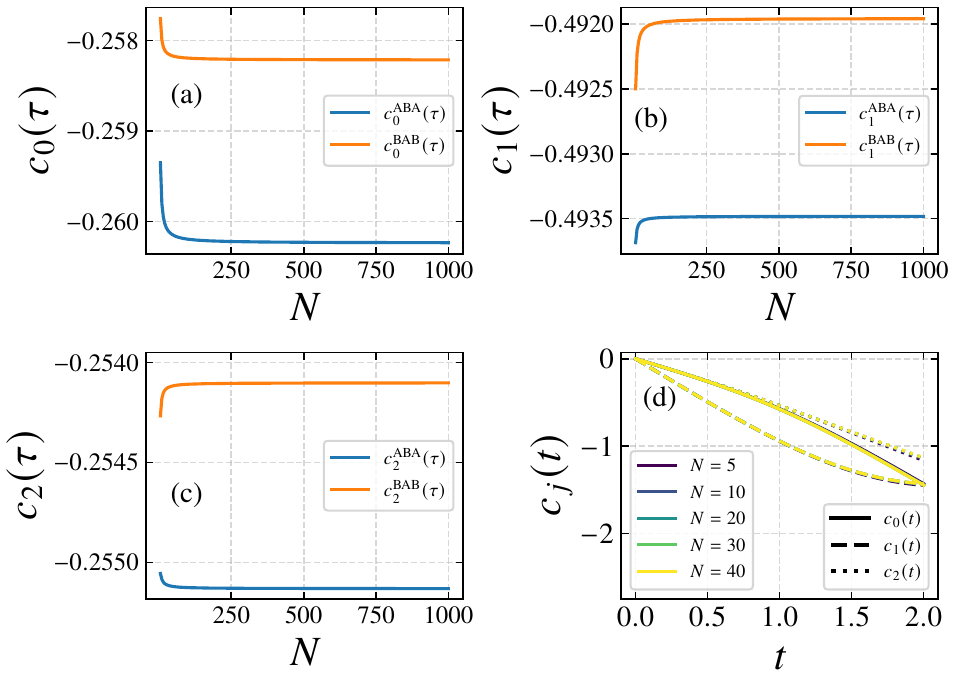}
    \caption{(a,b,c) The variational parameters for the QIM [Eq. \eqref{eqn:QIM}] with $h_x=h_z=J=1$ as a function of $N$ at time step $\tau=0.5$ up to 1000 qubits. (d) A plot of the variational parameters in the ansatz Eq. \eqref{eqn:ansatz_V2} for the QIM [Eq. \eqref{eqn:QIM}] with $h_x=h_z=J=1.0$ as a function of time $t$ for different number of qubits, $N$, as indicated. Here $t$ is in units of $1/J$.}
    \label{fig:Ising_varparm_vs_N}
\end{figure}

A natural question to ask is how the performance of the variational approach depends on the system size (i.e., the number of spins). In particular, solving Eq. \eqref{eqn:diff_eq_var_parms} requires computing traces involving matrix exponentials that could be computational expensive specially at large $N$. To address this concern, we considered the three-exponential and four-exponential ansatzes in Eqs.~\eqref{eqn:ansatz_V2} \& \eqref{eqn:ansatz_V3} and we have computed these traces analytically for arbitrary $N$ and coupling parameters, the results of which are shown in Appendix \ref{app:traces}. These expressions require no matrix exponentials and just simple function evaluations, resulting in a substantial reduction of computational cost to $\mathcal{O}(1)$. As an illustration, we analyzed the variational parameters appearing in Eq.~\eqref{eqn:ansatz_V1} for the Ising model with $h_x = h_z = J = 1.0$ at $\tau = 0.5$ as shown in Fig.~\ref{fig:Ising_varparm_vs_N} (a), (b), and (c), we plot the variational parameters $c_0(\tau)$, $c_1(\tau)$, and $c_2(\tau)$ at both operator orderings ($ABA$ and $BAB$) as a function of $N$ for up to a thousand qubits. Clearly, the parameters smoothly converge to well-defined values as $N \to \infty$. 

\begin{figure}
    \centering
    \includegraphics[width=\linewidth]{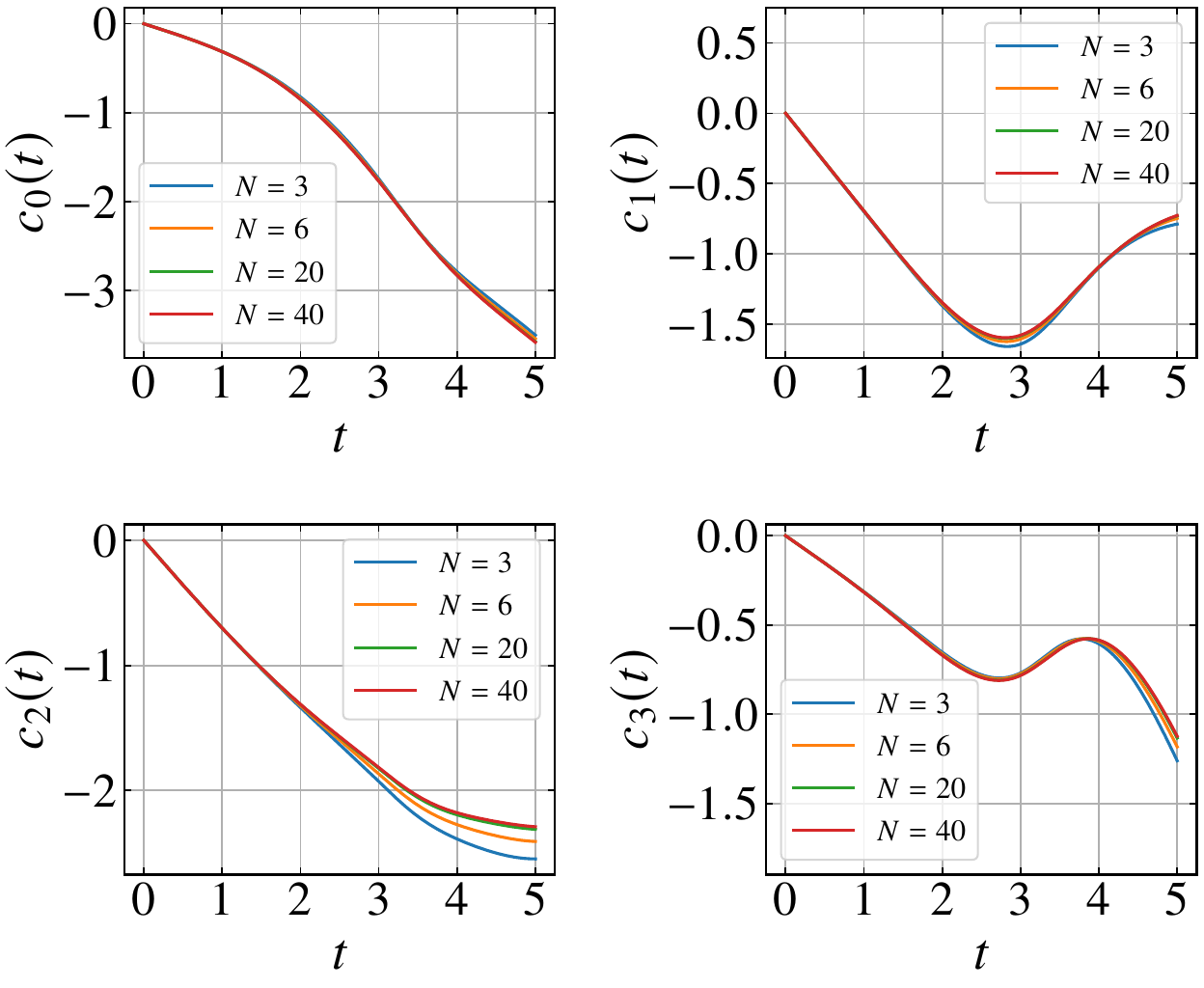}
    \caption{The variational parameters appearing in the unitary ansatz $U_a(t)=e^{ic_0(t)A}e^{ic_1(t)B}e^{ic_2(t)A}e^{ic_3(t)B}$ for the QIM [Eq. \eqref{eqn:QIM}] with $h_x=h_z=J=1$ different $N$ values as indicated. Here $t$ is in units of $1/J$.}
    \label{fig:Ising_varparm_vs_N_V4}
\end{figure}

We further examined the time dependence of the full set of variational parameters at different system sizes, shown in Fig.~\ref{fig:Ising_varparm_vs_N} (d). 
At short times, the parameters are largely insensitive to system size. 
At longer times, small deviations emerge, but the curves become increasingly closer as $N$ grows. 
This indicates convergence in the thermodynamic limit to well-defined functions $c_{j}^\infty(t)$, where
\begin{equation}
    c_{j}^\infty(t) = \lim_{N \to \infty} c_j(t).
\end{equation}
The same results were also found in the parameters appearing in the ansatz Eq. \eqref{eqn:ansatz_V3} as shown in Fig. \ref{fig:Ising_varparm_vs_N_V4}. For our Ising model we found that $g$ and $F$ have a common factor $2^N$ which cancels out in the equations of motion leaving only terms in both that grow at most linearly in $N$. Hence, the solutions to the EOM Eq. \eqref{eqn:diff_eq_var_parms}, namely $c_j(t)$, remain finite and well-behaved as $N\to\infty$. These numerical results are consistent with the analysis in Sec.~\ref{sec:approx_analytic_vars_AB}, where the approximate analytical formulas for the variational parameters were shown to admit a well-defined thermodynamic limit.


\begin{figure}
    \centering
    \includegraphics[width=\linewidth]{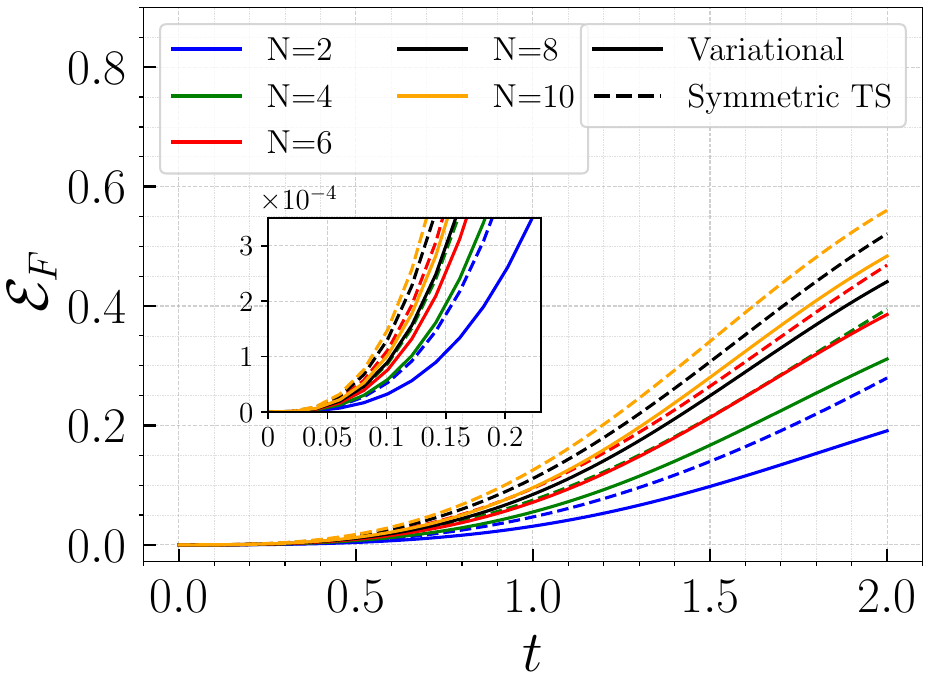}
    \caption{Normalized Frobenius error between the exact time-evolution operator and the approximate methods for the quantum Ising model [Eq.~\eqref{eqn:QIM}]. Results are shown for the variational ansatz [Eq.~\eqref{eqn:ansatz_V2}] and the symmetric Trotter–Suzuki formula [Eq.~\eqref{eqn:symmTS}] using the $AB$ operator ordering. Parameters are $J=h_z=h_x=1$. Here $t$ is in units of $1/J$.}
    \label{fig:errors_vs_N_Ising}
\end{figure}

Having established how the variational parameters $c_j(t)$ depend on the system size $N$, it is also instructive to examine how the global error $\mathcal{E}_F$ varies with $N$. To this end, we evaluate the normalized Frobenius error for the quantum Ising model at different system sizes and compare it with the corresponding error from the symmetric Trotter–Suzuki (TS) formula [Eq.~\eqref{eqn:symmTS}], as shown in Fig.~\ref{fig:errors_vs_N_Ising}. Clearly, the variational approach yields consistently lower errors across all system sizes, outperforming the equivalent TS decomposition even at small $t$, as highlighted in the inset of the figure. 

\subsection{The XXZ model}
\label{sec:xxz}

As a final example, we consider the anisotropic Heisenberg spin chain (the XXZ model), which plays a central role in the study of strongly correlated quantum systems \cite{XXZ_1,XXZ_2,XXZ_3}. 
This model exhibits rich physics ranging from gapless Luttinger-liquid behavior to gapped antiferromagnetic phases, and serves as a paradigmatic setting for exploring quantum magnetism, transport, and integrability \cite{XXZ_4,XXZ_5}. 
In addition, extensions of the XXZ chain with next-nearest-neighbor interactions are widely used to model frustration effects in low-dimensional materials \cite{XXZ_6,XXZ_7}. 

The XXZ Hamiltonian is
\begin{equation}
    H = H_{\rm NN} + H_{\rm NNN},
\end{equation}
where the nearest-neighbor (NN) and next-nearest-neighbor (NNN) contributions are given by
\begin{align}
    H_{\rm NN} &= J_1 \sum_{j=1}^{N-1} 
    \left( S_j^x S_{j+1}^x + S_j^y S_{j+1}^y + \Delta_1 S_j^z S_{j+1}^z \right), \\
    H_{\rm NNN} &= J_2 \sum_{j=1}^{N-2} 
    \left( S_j^x S_{j+2}^x + S_j^y S_{j+2}^y + \Delta_2 S_j^z S_{j+2}^z \right).
\end{align}
Here $S_j^{\alpha}$ ($\alpha = x,y,z$) are spin-$1/2$ operators on site $j$. 
The couplings $J_1$ and $J_2$ control the strength of the nearest- and next-nearest-neighbor exchange interactions, respectively, while $\Delta_1$ and $\Delta_2$ parameterize the anisotropy along the $z$ direction. 
For $\Delta_{1,2}=1$ the model reduces to the isotropic Heisenberg chain, and the case when $\Delta_{1,2}=0$ corresponds to the XX model. 

The next step is about choosing the time-evolution ansatz for this model. At first, we focus on the simple (integrable) case without NNN interactions, $J_2=0$. In this case the Hamiltonian can be conveniently decomposed as a sum of two non-commuting terms acting on alternating bonds,
\begin{align}
    A &= J_1 \sum_{j \ \text{even}}^{N-1} 
    \left( S_j^x S_{j+1}^x + S_j^y S_{j+1}^y + \Delta_1 S_j^z S_{j+1}^z \right), \label{eqn:XXZ_A}\\
    B &= J_1 \sum_{j \ \text{odd}}^{N-1} 
    \left( S_j^x S_{j+1}^x + S_j^y S_{j+1}^y + \Delta_1 S_j^z S_{j+1}^z \right)\label{eqn:XXZ_B}.
\end{align}
This even–odd splitting is especially useful for Trotter or variational decompositions, since all terms within $A$ ($B$) mutually commutes with each other, allowing for efficient implementation on both classical and quantum hardware.

\begin{figure}
\centering
    \includegraphics[width=0.99\linewidth]{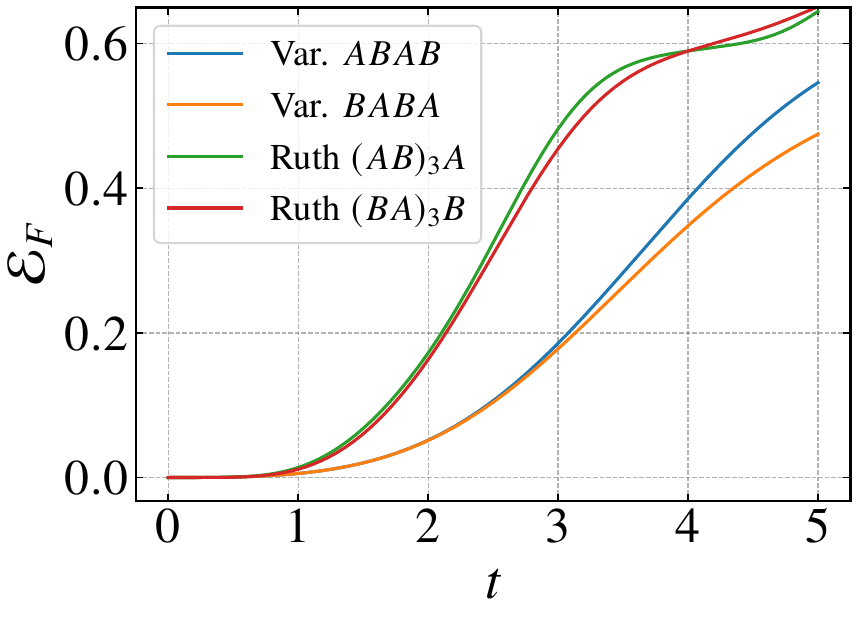}
\caption{Normalized Frobenius norm of the difference between exact and approximate time evolution operators for the XXZ model with only NN interaction, obtained using our four-exponential variational ansatz [Eq.~\eqref{eqn:ansatz_V3}] and Ruth's formula [Eq.~\eqref{eqn:ruth}] at different operator orderings. For the variational ansatz: ABAB and BABA; for Ruth: $(AB)_3A$ and $(BA)_3B$ (denoting ABABABA and BABABAB, respectively). Parameters: $J_1=1.0$, $\Delta_1=0.9$, $N=6$. Here $\tau$ is in units of $1/J_1$.}
\label{fig:XXZ_V4_frob}
\end{figure}

Figure~\ref{fig:XXZ_V4_frob} extends our analysis to the XXZ chain with parameters $J_1=1.0$, $\Delta_1=0.9$, and $N=6$, comparing the Frobenius norm error of our four-exponential variational ansatz (Eq. \eqref{eqn:ansatz_V3}) against Ruth's seven-exponential formula (Eq.~\eqref{eqn:ruth}). Remarkably, the variational approach maintains its superior accuracy despite requiring nearly half the number of exponentials per layer. This consistent performance across different physical models—both the quantum Ising and XXZ chains—demonstrates the robustness of our method. The observed reduction in circuit depth, coupled with enhanced precision, reinforces the variational ansatz as a versatile tool for efficient quantum simulations.

Now, we want to include the NNN interactions where we have $J_2 \neq 0$. Such a two-term splitting is no longer possible for this model. Instead, the Hamiltonian can be decomposed into into more than two parts. For example, one could choose $H = A + B + C$, where
\begin{equation}
\label{eqn:XXZ_A}
    A = J_1 \sum_{j=1}^{N-1} S_j^x S_{j+1}^x
      + J_2 \sum_{j=1}^{N-2} S_j^x S_{j+2}^x ,
\end{equation}
\begin{equation}
\label{eqn:XXZ_B}
    B = J_1 \sum_{j=1}^{N-1} S_j^y S_{j+1}^y
      + J_2 \sum_{j=1}^{N-2} S_j^y S_{j+2}^y ,
\end{equation}
\begin{equation}
\label{eqn:XXZ_C}
    C = J_1 \Delta_1 \sum_{j=1}^{N-1} S_j^z S_{j+1}^z
      + J_2 \Delta_2 \sum_{j=1}^{N-2} S_j^z S_{j+2}^z .
\end{equation}
At this stage, we proceed with one of two ways for constructing the variational ansatz either by using Eq. \eqref{eqn:ansatzABC} or Eqs. (\eqref{eqn:ansatzABC_2steps_1}-\eqref{eqn:ansatzABC_2steps_3}). The first is a single-step split (SSS), where we use Eq. \eqref{eqn:ansatzABC} and consider the ansatz,  
\begin{equation}
\label{eqn:XXZ_single_ansatz}
    U_a^{\rm SSS}(t)=e^{i c_{0}(t) A}\,e^{i c_{1}(t) B}\,e^{i c_{2}(t) C}\,e^{i c_{3}(t) B}\,e^{i c_{4}(t) A},
\end{equation}  

The second strategy is a two-step split (TSS). In the first step, we group $B$ and $C$ into a single block $\tilde{B}=B+C$ and employ the three-exponential ansatz of Eq.~\eqref{eqn:ansatz_V2},  
\begin{equation}
\label{eq:first_split}
    U_a^{\rm TSS}(t)=e^{i c_0(t) A}\,e^{i c_1(t)\tilde{B}}\,e^{i c_2(t) A}.
\end{equation}
In the second step, the exponential $e^{i c_1(t)\tilde{B}}$ is further decomposed using another three-exponential ansatz,  
\begin{equation}
\label{eq:second_split}
    U_1(t)=e^{i c_3(t) B}\,e^{i c_4(t) C}\,e^{i c_5(t) B},
\end{equation}
where $\tilde{B}$ plays the role of an effective Hamiltonian and $-c_1(t)$ acts as the evolution time. Together, we obtain the TSS ansatz $U_a^{\rm TSS}(t)=e^{i c_0(t) A}\,U_1(t)\,e^{i c_2(t) A}$.  

\begin{figure}
    \centering
    \includegraphics[width=\linewidth]{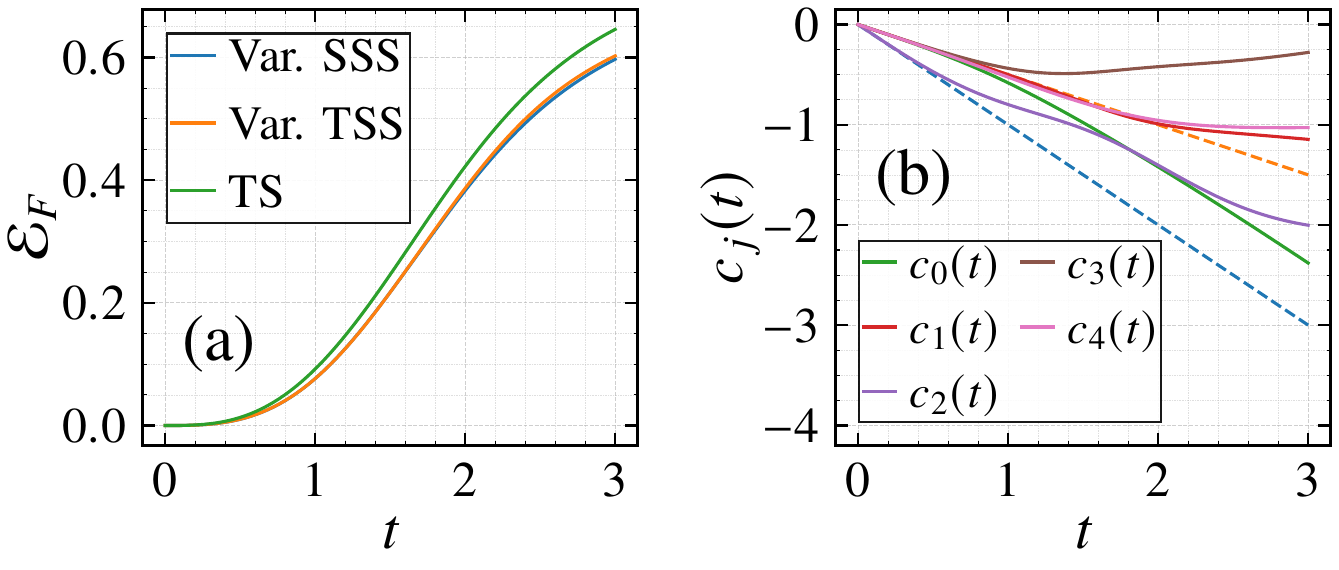}
    \caption{(a) A comparison of the Frobenius norm errors in the time-evolution obtained with the one-step split, two-step split, and symmetric Trotter--Suzuki decompositions for the $XXZ$ model with parameters $J_1=2.0$, $J_2=0.5$, $\Delta_1=\Delta_2=0.2$, and $N=5$. (b) Variational parameters extracted from the one-step split ansatz as functions of $t$. The dashed blue (orange) lines correspond to the Trotter--Suzuki parameters $-t$ ($-t/2$). Here $t$ is in units of $2/J_1$.}
    \label{fig:XXZ_split}
\end{figure}

To assess the performance of these approaches, we compare both ansatz with the symmetric Trotter--Suzuki formula for the $XXZ$ model with parameters $J_1=2.0$, $J_2=0.5$, $\Delta_1=\Delta_2=0.2$, and $N=5$ spins [see Fig.~\ref{fig:XXZ_split}(a)]. Both variational constructions yield a clear improvement over the Trotter--Suzuki results. For this parameter set, the single- and two-step splits exhibit nearly identical accuracy, with differences becoming apparent only at larger evolution times $t$ but this could change for other models. In general, however, the single-step split is expected to provide more robust performance across different Hamiltonian parameters as it is less constrained compared to the TSS ansatz.  

We have plotted the corresponding variational parameters obtained for the ansatz in Eq.~\eqref{eqn:XXZ_single_ansatz} are shown in Fig.~\ref{fig:XXZ_split}(b). At short times, the parameters follow the linear dependence prescribed by the Trotter--Suzuki formula (dashed lines). At longer times, nonlinear corrections generated by the variational procedure emerge, systematically enhancing the overall accuracy of the approximation.

To further benchmark our approach, we compare against a higher-order TS decomposition:
\begin{equation}
\label{eqn:7exp_suz}
U_{\rm TS}(t)=U_{\rm BC}(t)e^{-i t A}U_{\rm BC}(t),
\end{equation}
where
\begin{equation}
U_{\rm BC}(t)=e^{-\frac{i t}{4}C}
e^{-\frac{i t}{2}B}
e^{-\frac{i t}{4}C}.
\end{equation}
This TS ansatz contains seven exponentials. We construct a variational ansatz with identical structure and gate complexity:
\begin{equation}
\label{eqn:7exp_var}
U_a(t)=\prod_{j=0}^{6} e^{i c_j(t) A_j},
\end{equation}
with $A_0=A_2=A_4=A_6=C$, $A_1=A_5=B$, and $A_3=A$.

\begin{figure}
    \centering
    \includegraphics[width=\linewidth]
    {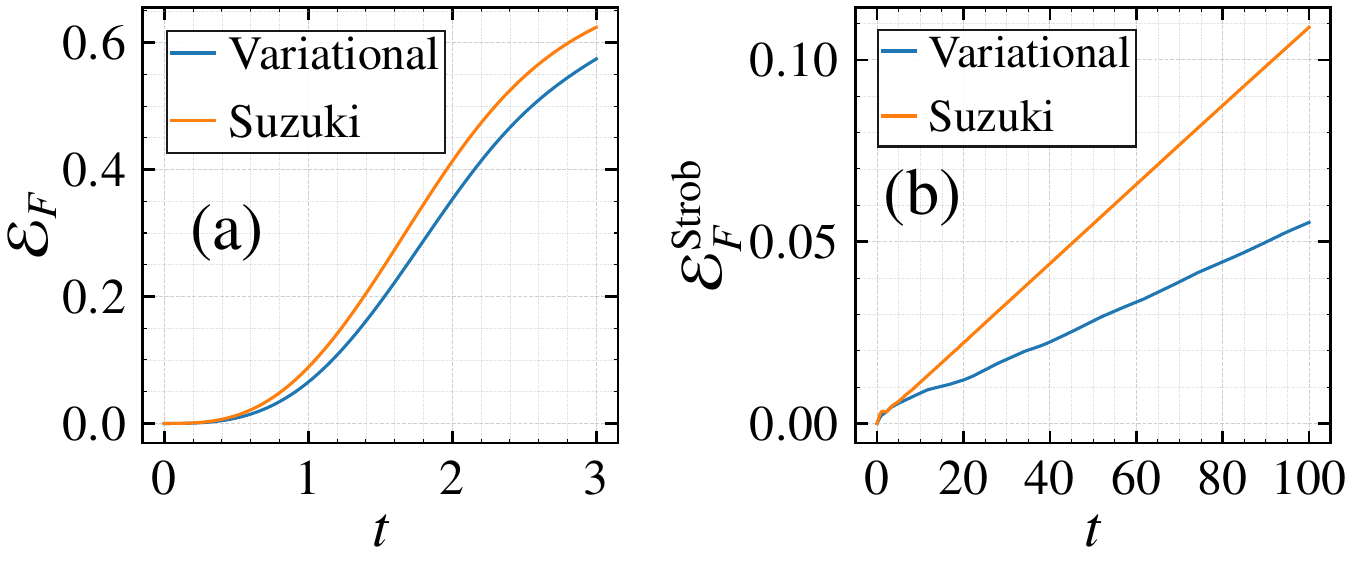}
    \caption{(a) A comparison of the Frobenius norm errors in the time-evolution obtained using the variational scheme of Eq.~\eqref{eqn:7exp_var} and the seven-exponential Trotter--Suzuki decomposition of Eq.~\eqref{eqn:7exp_suz} for the $XXZ$ model with parameters $J_1=2.0$, $J_2=0.5$, $\Delta_1=\Delta_2=0.2$, and $N=5$. (b) Stroboscopic error growth at longer simulation times, showing that the variational approach maintains better accuracy than the TS decomposition. Here $t$ is in units of $2/J_1$.}
    \label{fig:XXZ_7exp}
\end{figure}

Figure~\ref{fig:XXZ_7exp}(a) compares the normalized Frobenius norms, revealing that despite equivalent circuit depth, the variational approach achieves better accuracy. This advantage becomes particularly pronounced in long-time evolution, as shown in Fig.~\ref{fig:XXZ_7exp}(b) for $t=100$ with $\tau=0.2$. While TS error accumulates rapidly to approximately 11\% at $t=100$, the variational ansatz maintains substantially better accuracy ($\sim 5.5$\%)—representing a twofold improvement.

Nonetheless, higher-order ansatz can be employed to improve accuracy by following the same procedure outlined above. In addition, the Hamiltonian can be partitioned into more than three blocks if needed. For example, the nearest-neighbor (NN) terms may be split into two blocks, as discussed earlier, while the next-nearest-neighbor (NNN) terms can be grouped by interaction type, with the $XX$, $YY$, and $ZZ$ components each forming separate blocks. This yields a total of five blocks each with mutually commuting terms, for which a corresponding variational ansatz can be constructed to further enhance accuracy. Such extensions may prove useful for specific problems and are left for future investigation.  

\section{Building the Quantum Circuit for the Unitary Ansatz}
\label{sec:qc_impl}
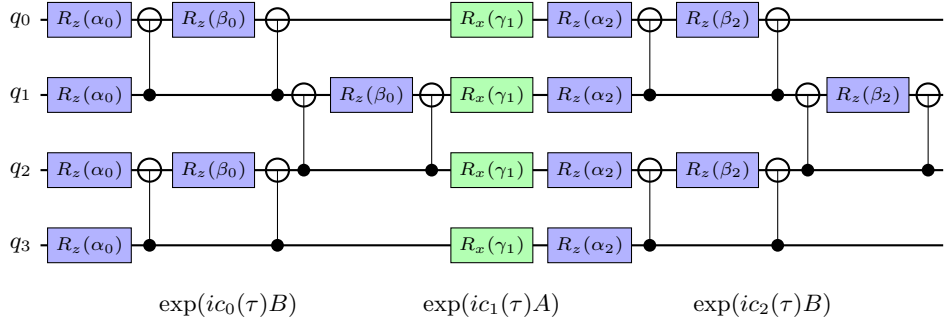
\begin{figure*}
\centering
\begin{adjustbox}{max width=\textwidth}
\begin{tikzpicture}[
    gate/.style={draw,minimum width=4mm,minimum height=4mm,fill=#1,text=black,font=\scriptsize},
    cnot/.style={fill=black,circle,inner sep=0.6mm},
    target/.style={draw,circle,minimum size=2mm,thick},
    line/.style={thick}
]
    \foreach \y/\pos in {0/3, 1/2, 2/1, 3/0} {
    \draw[line] (0,\pos) -- (12,\pos);
    \node[left] at (0,\pos) {$q_{\y}$};
}
    
    \foreach \y in {0,1,2,3} 
        \node[gate=blue!30] at (0.65,\y) {$R_z(\alpha_0)$};
    
    \node[cnot] at (1.45,0) {};
    \node[target] at (1.45,1) {};
    \draw (1.45,0) -- (1.45,1);
    
    \node[cnot] at (1.45,2) {};
    \node[target] at (1.45,3) {};
    \draw (1.45,2) -- (1.45,3);
    
    \node[gate=blue!30] at (2.3,1) {$R_z(\beta_0)$};
    \node[gate=blue!30] at (2.3,3) {$R_z(\beta_0)$};
    
    \node[cnot] at (3.15,0) {};
    \node[target] at (3.15,1) {};
    \draw (3.15,0) -- (3.15,1);
    
    \node[cnot] at (3.15,2) {};
    \node[target] at (3.15,3) {};
    \draw (3.15,2) -- (3.15,3);
    
    \node[cnot] at (3.5,1) {};
    \node[target] at (3.5,2) {};
    \draw (3.5,1) -- (3.5,2);
    
    \node[gate=blue!30] at (4.4,2) {$R_z(\beta_0)$};
    
    \node[cnot] at (5.2,1) {};
    \node[target] at (5.2,2) {};
    \draw (5.2,1) -- (5.2,2);
    
    \foreach \y in {0,1,2,3} 
        \node[gate=green!30] at (6.0,\y) {$R_x(\gamma_1)$};
    
    \foreach \y in {0,1,2,3} 
        \node[gate=blue!30] at (7.3,\y) {$R_z(\alpha_2)$};
    
    \node[cnot] at (8.1,0) {};
    \node[target] at (8.1,1) {};
    \draw (8.1,0) -- (8.1,1);
    
    \node[cnot] at (8.1,2) {};
    \node[target] at (8.1,3) {};
    \draw (8.1,2) -- (8.1,3);
    
    \node[gate=blue!30] at (9.0,1) {$R_z(\beta_2)$};
    \node[gate=blue!30] at (9.0,3) {$R_z(\beta_2)$};

    \node[cnot] at (9.8,0) {};
    \node[target] at (9.8,1) {};
    \draw (9.8,0) -- (9.8,1);
    
    \node[cnot] at (9.8,2) {};
    \node[target] at (9.8,3) {};
    \draw (9.8,2) -- (9.8,3);

    \node[cnot] at (10.2,1) {};
    \node[target] at (10.2,2) {};
    \draw (10.2,1) -- (10.2,2);
    
    \node[gate=blue!30] at (11,2) {$R_z(\beta_2)$};
    
    \node[cnot] at (11.8,1) {};
    \node[target] at (11.8,2) {};
    \draw (11.8,1) -- (11.8,2);
    
    \node at (2.5,-0.8) {$\exp(ic_0(\tau) B)$};
    \node at (6.0,-0.8) {$\exp(ic_1(\tau) A)$};
    \node at (9.6,-0.8) {$\exp(ic_2(\tau) B)$};
    
\end{tikzpicture}
\end{adjustbox}
\caption{Quantum circuit implementation of the time-evolution operator $U_a(\tau) = e^{i c_0(\tau) B} e^{i c_1(\tau) A} e^{i c_2(\tau) B}$ for the quantum Ising model [Eq. \eqref{eqn:QIM}] with $N=4$ qubits. Single-qubit rotations $R_x$ implement the transverse field, while longitudinal fields correspond to $R_z$ gates. The two-body $ZZ$ interactions are decomposed into CNOT–$R_z$–CNOT sequences. The gate angles are given by $\alpha_i = -2 c_i(\tau) h_z$, $\beta_i = -2 c_i(\tau) J$, and $\gamma_1 = -2 c_1(\tau) h_x$.}
\label{fig:ising_qc}
\end{figure*}

So far we have discussed the variational ansatz and its accuracy from a theoretical perspective. 
In practice, however, the usefulness of such an ansatz depends critically on how efficiently it can be implemented on quantum hardware. 
Here we demonstrate how the three-exponential ansatz in Eq.~\eqref{eqn:ansatz_V2} can be mapped to a quantum circuit for a standard case, the  quantum Ising model. 
For concreteness, we consider a chain of four qubits, as illustrated in Fig.~\ref{fig:ising_qc}.

The decomposition follows directly from the operator splitting introduced earlier, where we identify
\begin{equation}
    A = h_x \sum_{j=1}^N S_j^x, \qquad 
    B = J \sum_{j=1}^{N-1} S_j^z S_{j+1}^z + h_z \sum_{j=1}^N S_j^z.
\end{equation}
The $x$-field contribution is implemented as single-qubit $R_x$ rotations, while the $z$-field corresponds to $R_z$ rotations. 
The nearest neighbor (NN) $ZZ$ interactions are realized using a standard decomposition into a sequence of CNOT gates, a single $R_z$ rotation, and the corresponding inverse CNOT gates between every pair of neighboring qubits. 
This construction makes the ansatz directly compatible with existing superconducting and trapped-ion hardware, where such two-qubit gates are natively supported. 

As shown in Fig.~\ref{fig:ising_qc}, the operator operator 
\begin{equation}
    U_a(\tau) = e^{i c_0(\tau) B} e^{i c_1(\tau) A} e^{i c_2(\tau) B}
\end{equation}
is realized using layers of $R_x$ and $R_z$ gates interleaved with CNOT networks that generate the required NN $ZZ$ couplings, where one can use the approximate analytic expressions of $c_j(\tau)$ given in Eq. \eqref{eqn:3exp_approx_Ising} to find estimates for the elementary gate rotations. The circuit depth scales linearly with the number of qubits, and the number of variational parameters remains fixed, independent of system size. 
This feature highlights a key advantage of the variational ansatz: it achieves higher accuracy than a Trotter-Suzuki expansion at comparable or even reduced gate cost, while preserving an implementation that is straightforward on near-term devices.

In the next section, we introduce alternative approximate variational methods that can be implemented within our framework.

\section{Approximate variational techniques}
This section is organized into three parts. In the first two parts, we present an approach for deriving approximate analytical expressions for the variational parameters for Hamiltonians split into two and three blocks, respectively, and show how the leading correction to the well-known Trotter--Suzuki results naturally arises. This results in alternative renormalized product formulas that can be directly used in quantum simulations for any arbitrary Hamiltonian. The third part focuses on the use of partial-trace techniques for lowering computational expenses and their connection to wavefunction-based methods.

\subsection{Approximate Parameters for Hamiltonians Split into Two Blocks}
\label{sec:approx_analytic_vars_AB}

In this section, we will discuss how to derive approximate formulas of the variational parameters for Hamiltonians of the form $H=A+B$. The basic idea is that we want to expand the variational parameters in terms of Taylor series up to cubic orders to allow lowest corrections on top of the usual Trotter-Suzuki linear. Namely, we write $c_j(t)=\sum_{j=1}^{3}\frac{c_j^{(n)}(0)}{n!}t^n$, where $c_j^{(n)}(0)$ is the $n$th time derivative of $c_j(t)$ evaluated at $t=0$. Here we imposed the initial condition $c_j(0)=0$ which ensures that our ansatz set to the identity at $t=0$. The expansion coefficients, $c_j^{(n)}(0)$, are systematically determined through successive differentiation of the equations of motion. The first-order coefficients $c_j^{(1)}(0)$ are obtained by solving Eq.~\eqref{eqn:diff_eq_var_parms} at $t=0$. Higher-order coefficients follow from temporal derivatives of the equations of motion: differentiating Eq.~\eqref{eqn:diff_eq_var_parms} yields $c_j^{(2)}(0)$, and a second differentiation provides $c_j^{(3)}(0)$. This approach captures essential non-linear corrections beyond the conventional Trotter-Suzuki approximation while maintaining computational tractability through the recursive structure of the derivative hierarchy. One of course can go beyond cubic corrections if necessary following exactly the same steps outlined here. For the case of our simplest ansatz in Eq. \ref{eqn:ansatz_V1}, the variational parameters obtained via our first action Eq. \ref{eqn:action_S} are approximated to be

\begin{equation}
\label{eqn:taylor_cubic}
c_j(t)\approx-t+\frac{c_j^{(3)}(0)}{3!}t^3,
\end{equation}
where
\begin{equation}
\label{eqn:cubic_L1}
\begin{aligned}
     c_0^{(3)}(0)=-2\chi\mathrm{Tr}[AB],\quad 
       c_1^{(3)}(0)=2\chi\mathrm{Tr}[A^2],
\end{aligned}
\end{equation}
and
\begin{equation}
    \chi=\frac{\mathrm{Tr}[A^2B^2]-\mathrm{Tr}[(AB)^2]}{\mathrm{Tr}[A^2]\mathrm{Tr}[B^2]-(\mathrm{Tr}[AB])^2}.
\end{equation}
with the quadratic term in the expansion fully vanishes for arbitrary model Hamiltonians. However, using other action principles, for example the one corresponding to Eq. \eqref{eqn:action2_S} ($\mathcal{L}_2$), the cubic corrections are slightly modified to be
\begin{equation}
\begin{aligned}
    c_0^{(3)}(t)&\approx -2\left[\mathrm{Tr}[B^2]+\mathrm{Tr}[AB]\right]\chi\\
    c_1^{(3)}(t)&\approx 2\left[\mathrm{Tr}[A^2]+\mathrm{Tr}[AB]\right]\chi
    \end{aligned}.
\end{equation}
These cubic corrections are model specific and are new improvement on top the usual TS decomposition. For further analysis, we consider the three exponentials ansatz in Eq. \eqref{eqn:ansatz_V2}, both action principles $\mathcal{L}_1$ and $\mathcal{L}_2$ gave the same results below
\begin{equation}
\label{eqn:approx_cs_3exp}
\begin{aligned}
    c_{0}(t)&=c_{2}(t)\approx-\frac{t}{2}+\frac{c_{0,2}^{(3)}(0)}{3!}t^3,\\
    c_1(t)&\approx-t+\frac{c_1^{(3)}(0)}{3!}t^3,
\end{aligned}
\end{equation}
where
\begin{equation}
\begin{aligned}
    c_{0}^{(3)}(0)&=c_{2}^{(3)}(0)=-\chi\frac{\mathrm{Tr}[B^2]+\frac{1}{2}\mathrm{Tr}[AB]}{2},\\
    c_{1}^{(3)}(0)&=\chi\left[\mathrm{Tr}[AB]+\frac{1}{2}\mathrm{Tr}[A^2]\right].
\end{aligned}.
\end{equation}

The traces appearing in the above expressions can be computed analytically for the spin models at arbitrary coupling parameters and system sizes as illustrated in Appendix \ref{app:traces}. The interested reader can easily extend these ideas to higher order ansatz and obtain approximate formulas of the associated variational parameters. 

This greatly benefits quantum-hardware implementations, as the parameterized ansatz serves as a more accurate, Hamiltonian-specific alternative to Trotter--Suzuki decompositions. The variational parameters directly determine the rotation angles of the elementary gates, and the unitaries $e^{i c_0(\tau) B}$ and $e^{i c_1(\tau) A}$ are naturally realizable on quantum hardware because their exponentials factorize over the qubits. Moreover, the coefficients $c_j(\tau)$ are available in closed analytic form as functions of the system size $N$ and the Hamiltonian couplings, making them straightforward to compute for arbitrary models. This analytic accessibility ensures that the approach scales smoothly from few-qubit systems to the thermodynamic limit, providing a practical route toward accurate large-scale simulations on future quantum devices.

Another important use of these approximate formulas is the use in deciding what operator ordering to use. To clarify further, we start with Eq. \eqref{eqn:R_AB}, and use the approximation $\mathrm{Tr}[B\tilde{B}]\approx \mathrm{Tr}[B^2]+c_0^2\left(\mathrm{Tr}[(AB)^2]-\mathrm{Tr}[A^2B^2]\right)$, giving
\begin{align} 
\label{eqn:R_AB_app}
        R_{\rm AB}\approx&(\dot{c}_0+1)^2\mathrm{Tr}[A^2]+(\dot{c}_1+1)^2\mathrm{Tr}[B^2]\nonumber \\
        &2(\dot{c}_0+1)(\dot{c}_1+1)\mathrm{Tr}[AB]\nonumber\\
        &+2\dot{c}_1c_0^2\left(\mathrm{Tr}[(AB)^2]-\mathrm{Tr}[A^2B^2]\right)
\end{align}
Now, using Eqs. (\eqref{eqn:taylor_cubic} and \eqref{eqn:cubic_L1}), we find
\begin{align} 
\label{eqn:R_AB_quartic}
        R_{\rm AB}(t)\approx&-t^2\left(\mathrm{Tr}[(AB)^2]-\mathrm{Tr}[A^2B^2]\right)\nonumber \\
        &+\frac{\lambda}{3}\left(4\mathrm{Tr}[AB]-3\mathrm{Tr}[A^2]\right)t^4,
\end{align}
where
\begin{equation}
    \lambda=\chi\left[\mathrm{Tr}[A^2B^2]-\mathrm{Tr}[(AB)^2]\right]
\end{equation}
Thus, the ordering parameter becomes 
\begin{equation}
    \Delta={\rm sign}(\lambda\left(\mathrm{Tr}[B^2]-\mathrm{Tr}[A^2]\right))
\end{equation}
As an illustration, we consider our simple model in Eq. \eqref{eqn:2levelsH}, we find $\Delta={\rm sign}(h_z^2-h_x^2)$. Using $h_x=5$, $h_z=2$, we find $\Delta=-1$ which aligns with the earlier findings in Fig. \ref{fig:Two_Levels_Frob_Norm}(a) where the AB ordering gave smaller operator norm (i.e. more accurate than BA). Finding the ordering parameter for any other ansatz will follow the same steps outlined above. In the next section, we will find $\Delta$ for the QIM and the XXZ models.

In what follows, we will find the approximate variational parameters obtained via the first action principle for the spin models discussed previously.  

\subsubsection{Application to the Quantum Ising Model}
In this section, we will generate the approximate variational parameters for our transverse field Ising model obtained using the $\mathcal{L}_1$ action principle. First, for the ansatz in Eq. \eqref{eqn:ansatz_V1} with BAB operator ordering, we find
\begin{equation}
    c_0(t)\approx-t
\end{equation}
and
\begin{equation}
    c_1(t)\approx-t+\left[\frac{1}{12}\left(1-\frac{1}{N}\right)J^2+\frac{h_z^2}{6}\right]t^3
\end{equation}
where we have used the analytical traces obtained in Appendix \ref{app:traces}. Clearly, these approximate formulas have well-define thermodynamic limits which aligns with our earlier findings in Fig. \ref{fig:Ising_varparm_vs_N}. We further compute the ordering parameter for this ansatz to be
\begin{equation}
    \Delta=\operatorname{sign}\!\big(4N(h_x^2-h_z^2)-J^2(N-1)\big).
\end{equation}
where in the thermodynamic limit the AB operator ordering is the correct decomposition to proceed with whenever $h_x^2<h_z^2+\frac{J^2}{4}$, while we use the BA ordering when $h_x^2>h_z^2+\frac{J^2}{4}$.

Next, we find the variational parameters of our three exponentials ansatz in Eq. \eqref{eqn:ansatz_V2} to be
\begin{equation}
\label{eqn:3exp_approx_Ising}
\begin{aligned}
     c_0(t)&=c_2(t)\approx
-\frac{t}{2}-\frac{h_x^2}{12}\frac{J^2(N-1)+2h_z^2 N}{J^2 (N-1)+4h_z^2N}t^3\\
c_1(t)&\approx-t+\frac{1}{24}\left[\left(1-\frac{1}{N}\right)\frac{J^2}{2}+h_z^2\right]t^3
\end{aligned}.
\end{equation}

The linear terms in the above formulas agree with the well-known second order (symmetric) TS formula. The cubic corrections, which have well-defined thermodynamic limit, introduced here carry more information about the given Hamiltonian and are very useful when implementing stroboscopic time-evolutions as in Eq. \eqref{eqn:U_strop_general}. In the language of quantum circuits, these corrections modify the elementary gate rotations for a better accuracy than the TS counterpart.

\begin{figure}
    \centering
    \includegraphics[width=\linewidth]{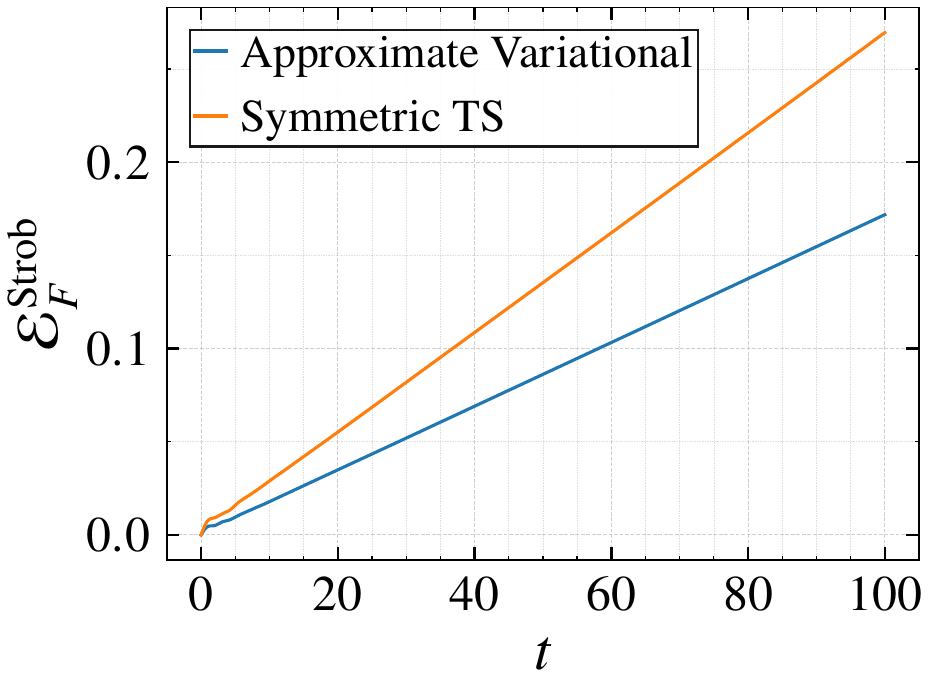}
    \caption{Stroboscopic error in the time evolution operator of the QIM[Eq.\eqref{eqn:QIM}] when comparing 
    the symmetric Trotter–Suzuki decomposition [Eq.~\eqref{eqn:symmTS}] 
    with the approximate variational approach [Eq.~\eqref{eqn:3exp_approx_Ising}]. 
    We set $h_x = h_z = J = 1.0$, $N=5$, and time step $\tau=0.1$. 
    The variational scheme exhibits a consistently smaller error growth with time. Here $t$ is in units of $1/J$.}
    \label{fig:approx_var_Ising}
\end{figure}

As an illustration, we benchmark the approximate variational ansatz of Eq.~\eqref{eqn:3exp_approx_Ising} against the symmetric Trotter--Suzuki decomposition of Eq.~\eqref{eqn:symmTS}. The stroboscopic error is shown in Fig.~\ref{fig:approx_var_Ising}. We observe that the variational construction yields a noticeably smaller error growth compared to the Trotter--Suzuki case, even for modest system sizes and time steps. This improvement highlights the advantage of tailoring the evolution operator with variational parameters, which can partially capture higher-order corrections that are absent in fixed Trotter decompositions. Further more, one can simply use the approximate variational parameters in Eq.~\eqref{eqn:3exp_approx_Ising} when computing any observable instead of directly employing the symmetric trotter-Suzuki formula.

\subsubsection{The Integrable XXZ model}
In this section, we will compute the approximate variational parameters for our XXZ model. For simplicity, we start with the integrable case where $J_2=0$. The relevant traces analytically in Appendix \ref{app:traces}. Using the simple two exponentials ansatz in Eq. \eqref{eqn:ansatz_V1}, we find
\begin{equation}
    c_0(t)\approx -t
\end{equation}
\begin{equation}
    c_1(t)\approx -t+\left[\frac{J_1^2}{12}\frac{1+2\Delta_1^2}{2+\Delta_1^2}\frac{N-2}{\left\lceil\frac{N-1}{2}\right\rceil}\right]t^3
\end{equation}
Interestingly, the ordering parameter for this example simplifies to $\Delta=\frac{1}{2}(1+(-1)^N)$ where the ordering matters when $N$ is even in which case the $BA$ ordering is preferred. We note that, unlike the Ising model, $\Delta$ is independent of the Hamiltonian parameters $J_1$ and $\Delta_1$, which has to do with the special operator splitting we chose for this model as given in Eqs. \eqref{eqn:XXZ_A} and \eqref{eqn:XXZ_B}. 

For the three exponentials ansatz (Eq. \eqref{eqn:ansatz_V2}), we find the variational parameters for our XXZ model to be
\begin{equation}
\begin{aligned}
     c_0(t)&=c_2(t)\approx -\frac{t}{2}-\frac{J_1^2}{48}\frac{1+2\Delta_1^2}{2+\Delta_1^2}\frac{N-2}{\left\lfloor\frac{N-1}{2}\right\rfloor}t^3\\
     c_1(t)&\approx -t+\frac{J_1^2}{48}\frac{1+2\Delta_1^2}{2+\Delta_1^2}\frac{N-2}{\left\lceil\frac{N-1}{2}\right\rceil}t^3.
\end{aligned}
\end{equation}

\subsection{Approximate Parameters for Hamiltonians Split into Three Blocks}
\label{sec:approx_analytic_vars_ABC}
In this section, we will discuss how to find approximate variational parameters for a general Hamiltonian $H=A+B+C=A+\tilde{B}$ based on the two step splitting discussed earlier. Basically, the first step begins with the splitting according to Eq. \eqref{eq:first_split}, and using the results from the previous section, we have 
\begin{equation}
c_j(t)\approx-t+\frac{c_j^{(3)}(0)}{3!}t^3,
\end{equation}
where
\begin{equation}
    c_{0}^{(3)}(0)=c_{2}^{(3)}(0)=-\frac{\mathrm{Tr}[\tilde{B}^2]+\frac{1}{2}\mathrm{Tr}[A\tilde{B}]}{2}\chi_1
\end{equation}
\begin{equation}
    c_{1}^{(3)}(0)=\left[\mathrm{Tr}[A\tilde{B}]+\frac{1}{2}\mathrm{Tr}[A^2]\right]\chi_1
\end{equation}
and
\begin{equation}
    \chi_1=\frac{\mathrm{Tr}[A^2\tilde{B}^2]-\mathrm{Tr}[(A\tilde{B})^2]}{\mathrm{Tr}[A^2]\mathrm{Tr}[\tilde{B}^2]-(\mathrm{Tr}[A\tilde{B}])^2}
\end{equation}
For the next step split Eq. \eqref{eq:second_split}, we obtain

\begin{equation}
c_{3,5}(t)\approx\frac{c_1(t)}{2}-\frac{c_{3,5}^{(3)}(0)}{3!}c_1^3(t)
\end{equation}
and
\begin{equation}
c_4(t)\approx c_1(t)-\frac{c_4^{(3)}(0)}{3!}c_1^3(t)
\end{equation}
where the cubic corrections are obtained again section \ref{sec:approx_analytic_vars_AB} results, giving
\begin{equation}
    c_{3}^{(3)}(0)=c_{5}^{(3)}(0)=-\frac{\mathrm{Tr}[C^2]+\frac{1}{2}\mathrm{Tr}[BC]}{2}\chi_2
\end{equation}

\begin{equation}
    c_{5}^{(3)}(0)=\left[\mathrm{Tr}[BC]+\frac{1}{2}\mathrm{Tr}[B^2]\right]\chi_2
\end{equation}
and
\begin{equation}
    \chi_2=\frac{\mathrm{Tr}[B^2C^2]-\mathrm{Tr}[(BC)^2]}{\mathrm{Tr}[B^2]\mathrm{Tr}[C^2]-(\mathrm{Tr}[BC])^2}
\end{equation}
These results can be easily applied to our XXZ model with NNN interactions. In fact, we have derived analytic forms of the traces appearing in these formulas for this model in Eqs. \eqref{eqn:firstXXZ_NNN_trace}-\eqref{eqn:lastXXZ_NNN_trace}.  

\subsection{Partial traces}
In quantum simulations, the physical system starts from a particular initial state $\ket{\psi_0}$ then the state evolves under the model Hamiltonian and one might be interested in looking at a dynamical observable like the magnetization in real time. Although our global approach still valid, one might get improvements by projecting out the irrelevant part of the Hilbert space using the Krylov subspace \cite{Kry1,Kry2}. For example, if we start with the actions in Eqs. \eqref{eqn:action_S} and \eqref{eqn:action2_S}, and we replace the full traces with partial traces over the Krylov basis states $\{\ket{K_j}\}$, in other words, we change $\mathrm{Tr}\left[\cdot\right]$ to $\mathrm{Tr}_{\mathcal{K}}\left[\cdot\right]$, where
\begin{equation}
\mathrm{Tr}_{\mathcal{K}}\left[\cdot\right]=\sum_j\bra{K_j}\cdot\ket{K_j}
\end{equation}
where the Krylov basis states are constructed from the finite vector space $\mathcal{K}=\{H^n\ket{\psi_0};n=0,1,\cdots,\mathcal{N}-1\}$, the corresponding equations of motion will be exactly similar to the previous ones except that the trace replaced with partial traces. Typically, one expects this to improve results over short period of times for smaller basis. It is also obvious that the partial traces have the advantage of being computationally less expensive than the full trace computations making it useful in certain applications.  

\begin{figure}
\centering
    \includegraphics[width=0.49\linewidth]{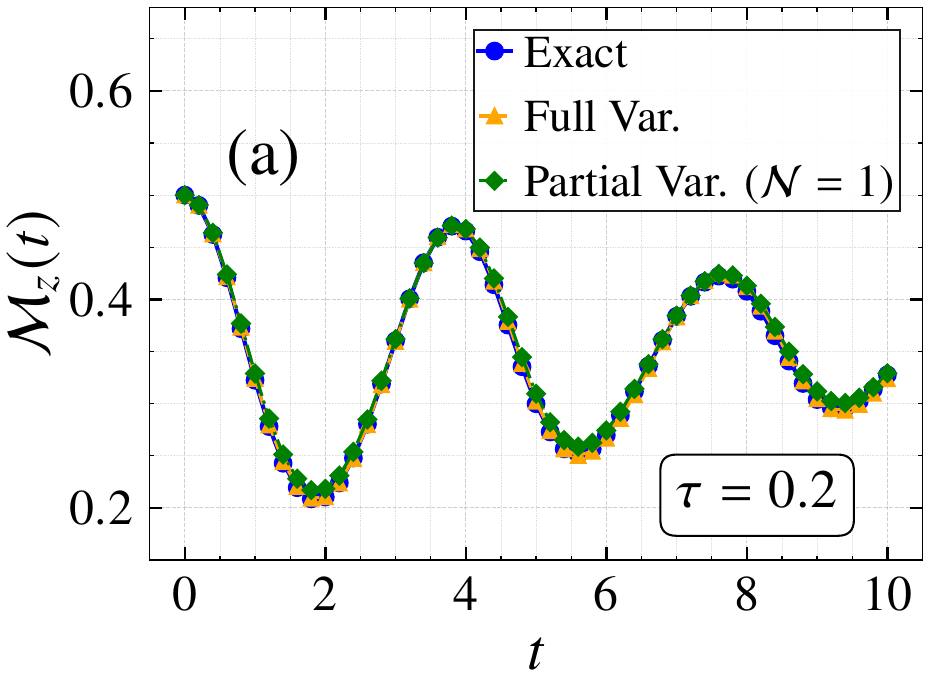}\hspace{0.005mm}
    \includegraphics[width=0.49\linewidth]{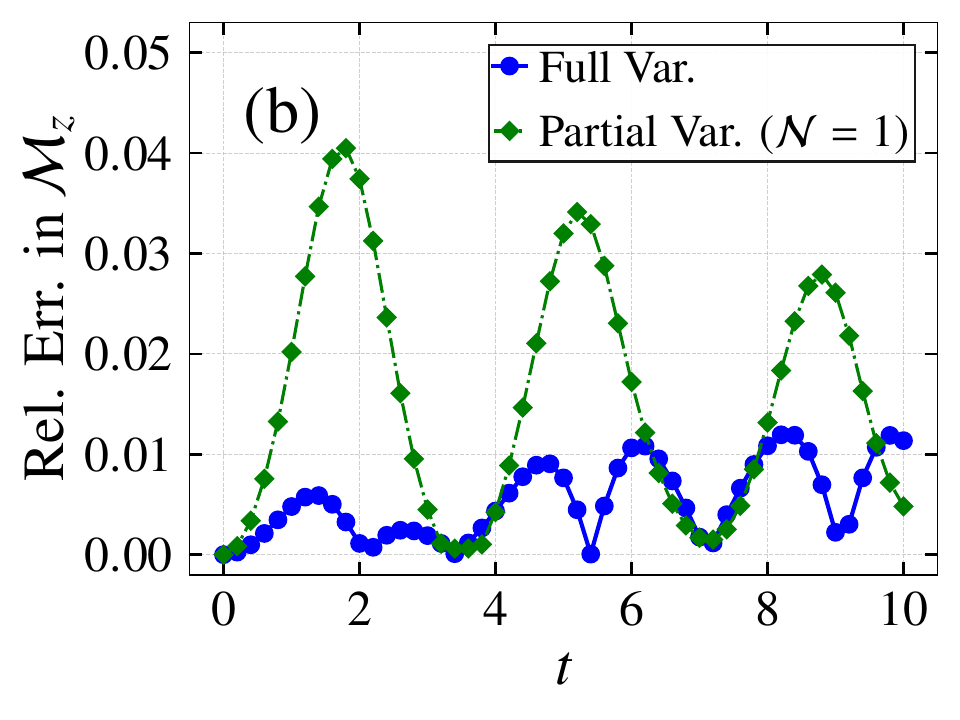}
    \hspace{0mm}
    \includegraphics[width=0.49\linewidth]{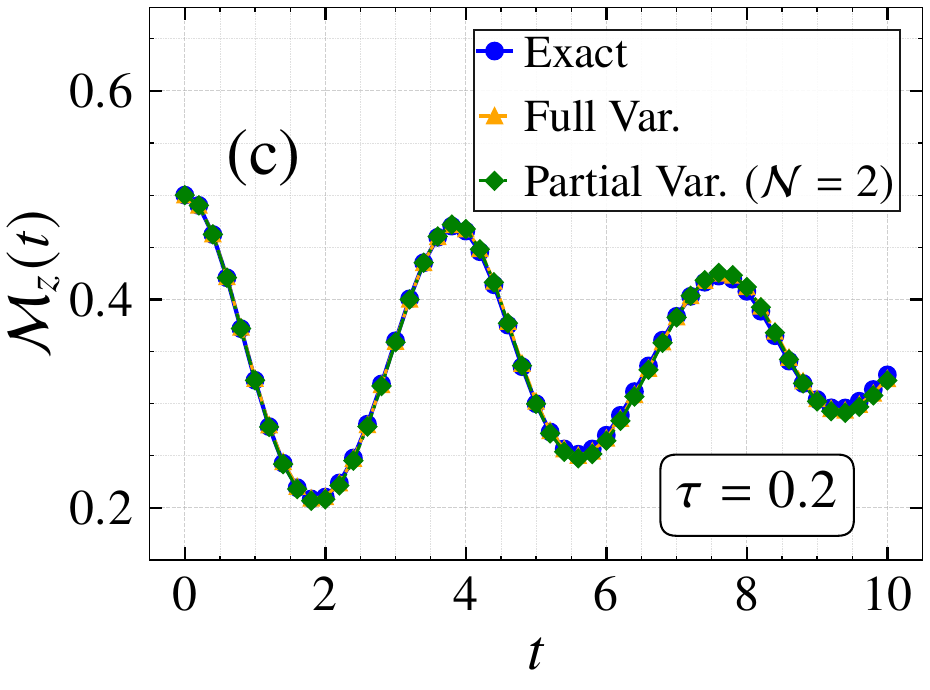}
    \includegraphics[width=0.49\linewidth]{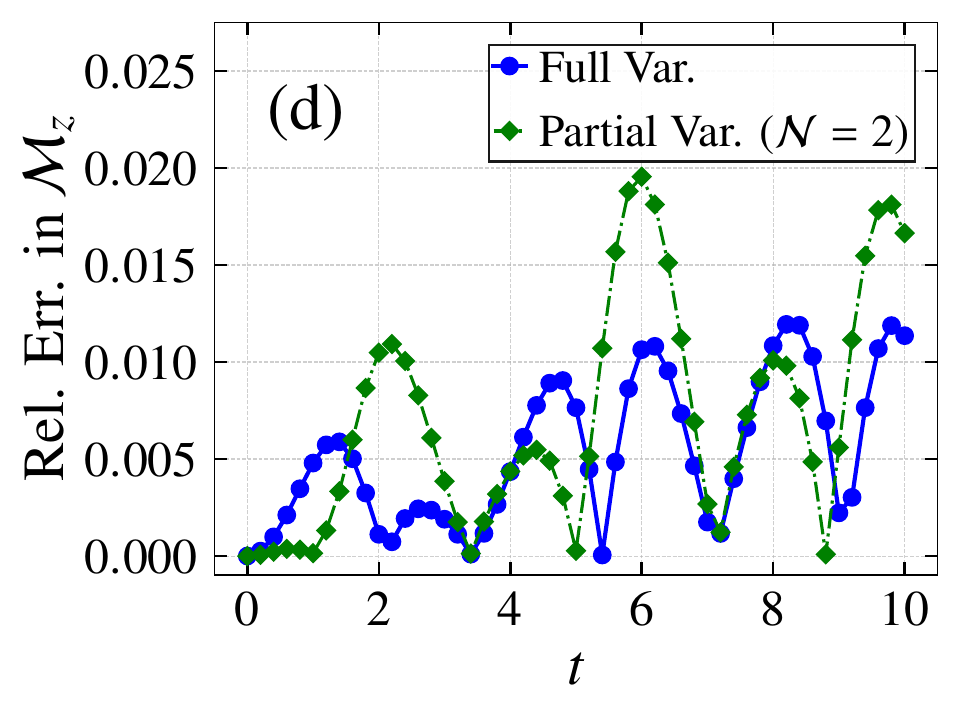}
    \hspace{0mm}
    \includegraphics[width=0.49\linewidth]{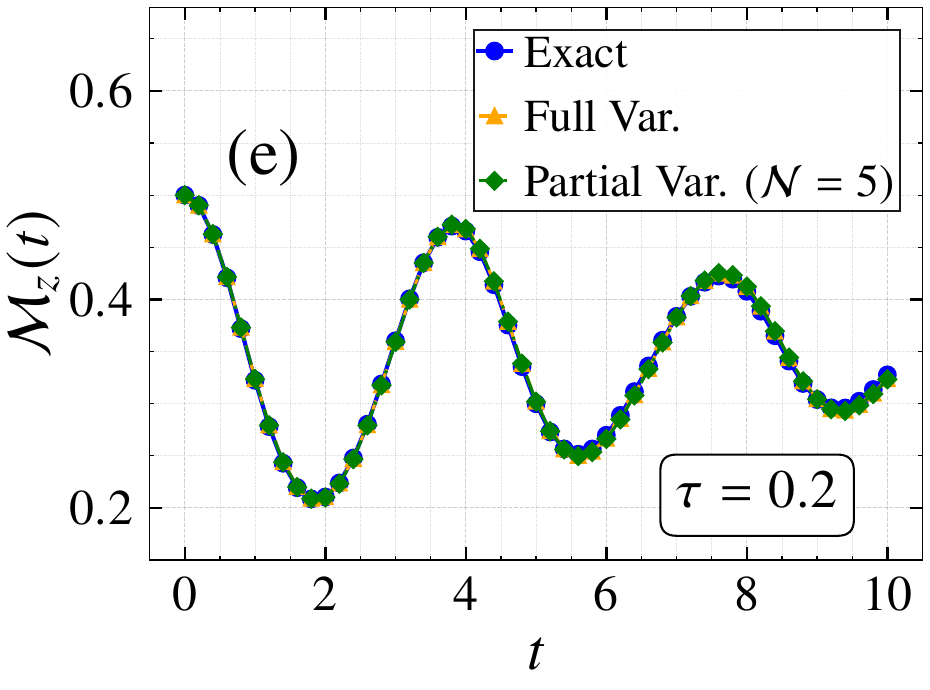}
    \includegraphics[width=0.49\linewidth]{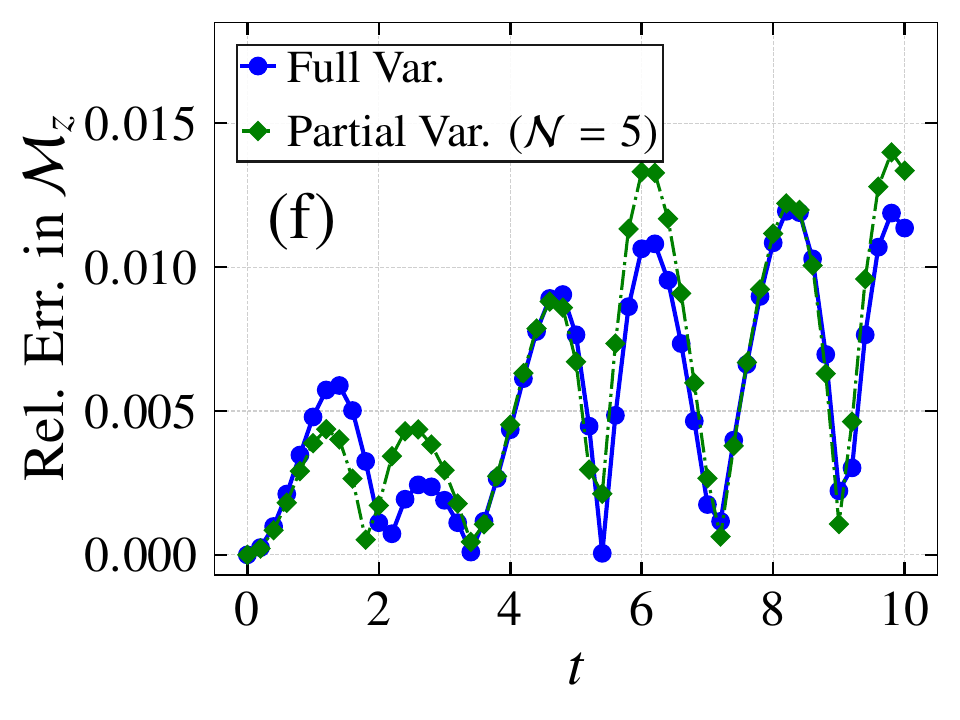}
\caption{The stroboscopic magnetization of the quantum Ising model  [Eq. \eqref{eqn:QIM}] obtained via full and partial traces variational settings with $h_x=h_z=J=1.0$ and $N=5$, using (a) one ($\mathcal{N}=1$), (c) two ($\mathcal{N}=2$), and (e) five ($\mathcal{N}=5$) Krylov basis vectors. (b,d,f) The corresponding relative errors in the dynamical magnetization. Here $t$ is in units of $1/J$.}
\label{fig:Ising_Krylov}
\end{figure}

As an illustration, we employed the idea of partial traces to compute the dynamical magnetization of the transverse-field Ising model with the ansatz in Eq. \eqref{eqn:ansatz_V2} and the parameters $h_x=h_z=J=1.0$ and $N=5$ as shown in Fig. \ref{fig:Ising_Krylov}. We have considered the cases with Krylov subspace of containing one, two, and five basis vectors and we compared those results against the full variational method as shown in Fig. \ref{fig:Ising_Krylov}(a,c,e). Moreover, we plotted the relative error in the magnetization in Fig. \ref{fig:Ising_Krylov}(b,d,f). We note that using 1 Krylov state, the partial trace results where less accurate than the full variational approach over the full simulation period as clear in \ref{fig:Ising_Krylov}(a). However, using few more Krylov vectors, we obtained better results than the full trace method only over the first few periods as shown in \ref{fig:Ising_Krylov}(d,f). This is because we have constrained our variational approach to a smaller region in Hilbert space that is relevant to the initial state which we are evolving.  

\section{Conclusions}
In this work, we introduced a variational product formula for approximating the full time evolution of arbitrary Hamiltonians, with variational parameters determined through a globally valid variational principle \cite{vogl2025variational}. Our ansatz is not only manifestly unitary but also structured as a product of elementary exponentials of local commuting terms, allowing it to be mapped directly onto standard quantum gate operations. We demonstrated that, across a range of scenarios, this approach achieves substantial improvements over standard Trotter–Suzuki decompositions. Moreover, we illustrated the flexibility of the method by applying it to systems ranging from simple models to complex many-body Hamiltonians.

We have also shown that our unitary variational ansatz can achieve higher accuracy with lower gate complexity compared to higher-order Trotter–Suzuki formulas such as the Ruth decomposition. In particular, our ansatz with four alternating exponentials outperformed the seven-exponential sequence of the Ruth formula. This highlights the strength of the renormalized global variational principles, which can be tailored to mimic realistic error environments when running simulations on quantum hardware. By explicitly accounting for coherence times, one can design ansätze that are better suited to the hardware constraints. An alternative perspective is to start from a custom-designed quantum circuit composed of a sequence of quantum gates, where the gate rotation angles serve as variational parameters. These parameters can then be optimized on a classical computer using our variational principle and subsequently deployed in quantum simulations on actual devices.

In addition, we briefly discuss several approximate variational techniques, including approximate analytical expressions for the variational parameters which provides renormalized product formulas that give improvements over the equivalent TS decomposition, and a partial-trace method inspired by the Krylov subspace approach. These methods are especially useful in settings where a fully exact treatment is either computationally infeasible or unnecessary.  

Our approach offers several notable advantages. First, it provides a stroboscopic approximation of $U(t)$ that remains accurate over longer time intervals than the Trotter–Suzuki method, which is primarily reliable only at very short times. Unlike state-specific variational techniques, the resulting unitary is state-independent and therefore applicable to \emph{any} initial state, making it a versatile tool for general-purpose quantum simulation. In addition, the ansatz typically achieves reduced circuit depth while delivering higher accuracy than Trotter–Suzuki formulas of comparable size—a key advantage in the NISQ era. We showed that for some Hamiltonians, our variational approach will provide accuracy superior to higher order TS decompositions. Beyond simulating dynamics, the framework also extends naturally to tasks such as constructing custom, hardware-efficient quantum gates \cite{custom_gates1, custom_gates2}. This is particularly valuable since high-fidelity two-qubit gates remain a major bottleneck in current quantum devices \cite{Two_QG_1, Two_QG_2}. By enabling the design of specialized circuit ansätze that minimize two-qubit interactions while preserving accuracy, our method directly addresses one of the central challenges in practical quantum computing.

We believe that our findings open promising avenues for real-world applications in quantum computing, with the potential to play a significant role in error mitigation—a cornerstone for advancing the accuracy and practicality of current quantum simulations.

\section{Acknowledgments}
 IA and JPFL acknowledges the support of the Natural Sciences and Engineering Research Council of Canada (NSERC) RGPIN-2022-03882 and (NRC) AQC-200-1. MK acknowledges the support from the Applied
Quantum Computing Challenge Program at the National Research Council of Canada.

\appendix
\begin{widetext}

\section{Equations of motion for variational parameters}
\label{app:EOM}
In this section, we will use Eq. \eqref{eqn:diff_eq_var_parms} to derive the explicit equations of motion for the three and four exponential ansatzes Eqs. \eqref{eqn:ansatz_V2} and \eqref{eqn:ansatz_V3}. 

\subsection{Three exponentials ansatz}
Consider the following time-evolution ansatz that approximates the exact time evolution of a model Hamiltonian $H$ 
\begin{equation}
    U_a(t)= e^{i c_0(t) A} e^{i c_1(t) B} e^{i c_2(t) A},
    \label{eq:3expAnsatz}
\end{equation}  
where \( A \), and \( B \) are Hermitian operators built from terms in $H=A+B$. The explicit equations of motion for the variational parameters are 
\begin{align}
    g_{00} \Dot{c_0} + g_{01} \Dot{c_1} + g_{02} \Dot{c_2} = D_0  \\
    g_{10} \Dot{c_0} + g_{11} \Dot{c_1} + g_{12} \Dot{c_2} = D_1  \\g_{20} \Dot{c_0} + g_{21} \Dot{c_1} + g_{22} \Dot{c_2} = D_2  
\end{align}
where $g_{ij}$ are the components of the quantum geometric tensor $g$
\begin{equation}
    g=\begin{bmatrix}
        \mathrm{Tr}[A^2]&\mathrm{Tr}[AB]& T_1(c_1)\\
        \mathrm{Tr}[AB]&\mathrm{Tr}[B^2]& \mathrm{Tr}[AB]\\
        T_1(c_1)&\mathrm{Tr}[AB]& \mathrm{Tr}[A^2]\\
    \end{bmatrix}
\end{equation}
where
\begin{equation}
    T_1(c_1)=\mathrm{Tr}\left[Ae^{ic_1B}Ae^{-ic_1B}\right]
\end{equation}
and,
\begin{equation}
\begin{aligned}
D_0 &= -\mathrm{Tr} \left[A^2\right]-\mathrm{Tr} \left[AB\right], \\
D_1 &= -\mathrm{Tr} \left[AB\right]-\mathrm{Tr} \left[Be^{ic_0A}Be^{-ic_0A}\right], \\
D_2 &= -T_1(c_1)-\mathrm{Tr} \left[e^{ic_1B}Ae^{-ic_1B}e^{-ic_0A}Be^{ic_0A}\right],
\end{aligned}
\end{equation}
with
\begin{equation}
    T_2(c_0)=\mathrm{Tr} \left[Be^{ic_0A}Be^{-ic_0A}\right].
\end{equation}
and
\begin{equation}
    T_3(c_0,c_1)=\mathrm{Tr} \left[e^{-ic_0A}Be^{ic_0A}e^{ic_1B}Ae^{-ic_1B}\right].
\end{equation}
These traces appearing in the EOMs can be computed analytically for general multi-qubit models as illustrated in Appendix \ref{app:traces}. Also, it should be noted that one may consider the other operator ordering where the ansatz is $U_0(t)=e^{ic_0(t)B}e^{ic_1(t)A}e^{ic_2(t)B}$, in which case the equations of motion are obtained from the above equations by simple swap $A\leftrightarrow B$.

\subsection{Four exponentials ansatz}
We further consider the following unitary ansatz 
\begin{equation}
    U_a(t)= e^{i c_0(t) A} e^{i c_1(t) B} e^{i c_2(t) A} e^{i c_3(t) B},
    \label{eq:4expAnstaz}
\end{equation}  

The corresponding EOMs for the variational parameters are
\begin{align}
    g_{00} \Dot{c_0} + g_{01} \Dot{c_1} + g_{02} \Dot{c_2}+ g_{03} \Dot{c_3} = D_0  \\
    g_{10} \Dot{c_0} + g_{11} \Dot{c_1} + g_{12} \Dot{c_2}+ g_{13} \Dot{c_3} = D_1  \\
    g_{20} \Dot{c_0} + g_{21} \Dot{c_1} + g_{22} \Dot{c_2}+ g_{23} \Dot{c_3} = D_2\\ 
    g_{30} \Dot{c_0} + g_{31} \Dot{c_1} + g_{32} \Dot{c_2}+ g_{33} \Dot{c_3} = D_3
\end{align}

where $g_{ij}$ are the components of the QGT
\begin{equation}
    g=\begin{bmatrix}
        \mathrm{Tr}[A^2]&\mathrm{Tr}[AB]& T_1(c_1)&T_3(-c_2,-c_1)\\
        \mathrm{Tr}[AB]&\mathrm{Tr}[B^2]& \mathrm{Tr}[AB]&T_2(c_2)\\
        T_1(c_1)&\mathrm{Tr}[AB]& \mathrm{Tr}[A^2]&\mathrm{Tr}[AB]\\
        T_3(-c_2,-c_1)&T_2(c_2)& \mathrm{Tr}[AB]&\mathrm{Tr}[B^2]
    \end{bmatrix}
\end{equation} 
and $D_j$ are the entries of the following vector
\begin{equation}
    D=-\begin{bmatrix}
      \mathrm{Tr}[AB]+\mathrm{Tr}[A^2]\\
      \mathrm{Tr}[AB]+T_2(c_0)\\
      T_1(c_1)+T_3(c_0,c_1)\\
      T_3(-c_2,-c_1)+T_4(c_0,c_1,c_2)
    \end{bmatrix},
\end{equation}
with 
\begin{equation}
T_4(c_0,c_1,c_2)=\mathrm{Tr}\left[e^{ic_2A} B e^{-ic_2A} e^{-ic_1B} e^{-ic_0A} B e^{ic_0A} e^{ic_1B}  \right]
\end{equation} 
Where the static and dynamical traces appearing above have been calculated analytically for the Ising model as shown in Appendix \ref{app:traces}. For completeness, the other operator ordering ansatz $U_a(t)= e^{i c_0(t) B} e^{i c_1(t) A} e^{i c_2(t) B} e^{i c_3(t) A}$ can be obtained by simple replacement $A\leftrightarrow B$ in the EOMs. However, it gives rise to new dynamical trace of the form
\begin{equation}
T_5(c_0,c_1,c_2)=\mathrm{Tr}\left[e^{ic_2B} A e^{-ic_2B} e^{-ic_1A} e^{-ic_0B} A e^{ic_0B} e^{ic_1A}  \right].
\end{equation}
which was also computed analytically for the QIM in Appendix \ref{app:traces}. 

\section{Alternative action principles}
\label{sec:alternative_S}
In this section, we will illustrate few examples of Lagrangians that can be used as basis for our variational principle. The first example we would like to consider is the following Lagrangian \cite{vogl2025variational}
\begin{equation}
\label{eqn:action2_S}
    \mathcal{L}_2=||i\partial_tU-HU||_F^2
\end{equation}
where $||X||_F=\sqrt{\, \mathrm{Tr} \left[XX^\dagger\right]}$ is the Frobenius norm of $X$. One readily finds $\partial_t^2 U+H^2U=0$ as variational equations of motion. Of course this is the time derivative of equation \eqref{eqn:SchEq} and therefore yields identical results if we choose $\dot{U}(0)=-iHU(0)$ as extra initial condition. The later constraint is very important as the second order differential equation have a general solution $U(t)=e^{-iHt}U_1+e^{iHt}U_2$ and the constraints forces $U_2=0$ and we recover the physical time-evolution. 

For a time-evolution ansatz $U_a$ expressed in terms of variational parameters $\{c_i(t)\}$, we obtain the following equations of motion
\begin{equation}
    \sum_k g_{jk}\ddot{c}_k+\sum_{\ell k}\Gamma_{j,\ell k}\dot{c}_\ell\dot{c}_k+2i\sum_k \mathcal{B}_{jk}\dot{c}_k+\mathcal{F}_j=0
\end{equation}
where $g$ is the quantum metric that appeared earlier in Eq. \eqref{eqn:diff_eq_var_parms}, $\Gamma_{j,\ell k}=\mathrm{Tr}\left[\left( \frac{\partial U_a}{\partial c_j} \right)^\dagger \frac{\partial^2 U_a}{\partial c_\ell \partial c_k}\right]$ is a Christoffel-like symbol derived from the metric $g$ which is classically analogous to the gradient of mass tensor in case of coupled Harmonic oscillators with position dependent effective mass, $\mathcal{B}_{jk}=\mathrm{Tr}\left[\left( \frac{\partial U_a}{\partial c_j} \right)^\dagger H\frac{\partial U_a}{\partial c_k}\right]$ measures how the expectation value of the Hamiltonian (up to a constant) varies with respect to parameters changes throughout the entire Hilbert space which is classically similar to the pseudo-magnetic field that represents variations in the vector potential, and $\mathcal{F}_j=-\mathrm{Tr} \left[ \left( \frac{\partial U_a}{\partial c_j} \right)^\dagger H^2 U_a \right]$ is a generalized force derived from a potential $V(c,c^*)=-\mathrm{Tr}\left[U^\dagger_a H^2 U_a\right]$, i.e. $\mathcal{F}_j=\partial_{c^*}V(c,c^*)$.

Although these equations of motion have richer structure compared to the first order differential equations associated with the first variational principle, they have the disadvantage of being second-order non-linear differential equations that are computationally more expensive but still feasible.     

In addition, one can also try the following Lagrangian
\begin{equation}
    \mathcal{L}_3=\frac{1}{2}\mathrm{Tr}\left[\dot{U}^\dagger \dot{U}-U^\dagger H^2 U\right]
\end{equation}
where the first term is similar to the kinetic energy term for a system of harmonic oscillator and the second term is the just the coupling potential energy with $H^2$ plays a rule of generalized spring constant matrix. In this case, the equations of motion are

\begin{equation}
    \sum_k g_{jk}\ddot{c}_k+\sum_{\ell k}\Gamma_{j,\ell k}\dot{c}_\ell\dot{c}_k-\mathcal{F}_j=0
\end{equation}

which are identical to the earlier Lagrangian but this time without the linear \emph{velocity} term.

One can even use more constrained Lagrangian
\begin{equation}
    \mathcal{L}_4=\mathrm{Tr}\left[\Lambda^\dagger X+X^\dagger \Lambda\right]
\end{equation}
where $X=i\dot{U}-HU$ is the residual operator. However, we will not exhaust all the possibilities in this work and are left as exercise for the interested reader.  
\section{Analytic calculations of the traces}
\label{app:traces}
\subsection{Static Traces}
Static traces appeared in the equations of motion (see Appendix \ref{app:EOM}) as well as in the approximate analytic expressions of the variational parameters derived in sections \ref{sec:approx_analytic_vars_AB} and \ref{sec:approx_analytic_vars_ABC}. In this section, we will derive analytic expressions for those traces for the Ising model and the XXZ model with NNN interaction. The idea is based on two facts. First, A spin operator at site $j$ can be expressed as tensor product
\begin{equation}
    \sigma_j^\alpha=I\otimes I\otimes\cdots \otimes \sigma^\alpha\otimes I\otimes\cdots\otimes I
\end{equation}
where $\alpha=x,y,z$ and $\sigma^\alpha$ is the $2\times2$ Pauli matrix. The other identity we will use the following
\begin{equation}
    \mathrm{Tr}[X\otimes Y]=
    \mathrm{Tr}[X]\mathrm{Tr}[Y]
\end{equation}
For the transverse field Ising model, we have $A=\frac{J}{4}\sum_{j=1}\sigma_j^z\sigma_{j+1}^z+\frac{h_z}{2}\sum_{j=1}^N\sigma_j^z$ and $B=\frac{h_x}{2}\sum_{j=1}^N\sigma_j^x$. First, one can easily see that $\mathrm{Tr}[AB]=0$. Next, let's look at the expressions of $A^2$ and $B^2$
\begin{equation}
A^2=\frac{J^2}{16}\sum_{j,m=1}^{N-1}\sigma_j^z\sigma_{j+1}^z\sigma_m^z\sigma_{m+1}^z+\frac{Jh_z}{4}\sum_{j=1}^{N-1}\sum_{n=1}^N\sigma_j^z\sigma_{j+1}^z\sigma_n^z+\frac{h_z^2}{4}\sum_{\ell,n=1}^N\sigma_{\ell}^z\sigma_n^z
\end{equation}
\begin{equation}
B^2=\frac{h_x^2}{4}\sum_{\ell,n=1}^N\sigma_{\ell}^x\sigma_n^x
\end{equation}
One can easily verify that $\mathrm{Tr}[B^2]=\frac{N 2^Nh_x^2}{4}$. Now, for $A^2$, let's walk through term by term. For the first term, we expect that non-zero contributions to the trace to come from the case when $j=m$, giving a total contribution of $\frac{(N-1)2^NJ^2}{16}$. Furthermore, the second term in $A^2$ gives zero contribution to the trace. Finally, the last term contributes $\frac{N2^Nh_z^2}{4}$ to the trace. This gives
\begin{equation}
    \mathrm{Tr}[A^2]=\frac{(N-1)2^NJ^2}{16}+\frac{N2^Nh_z^2}{4}
\end{equation}
Now, there are two more traces to compute. First, let's look at $A^2B^2$
\begin{equation}
A^2B^2=\frac{Nh_x^2}{4}A^2+\frac{1}{2}h_x^2A^2\sum_{i<j}\sigma_i^x\sigma_j^x
\end{equation}
which gives $\mathrm{Tr}[A^2B^2]=\frac{N2^Nh_x^2}{4}\left[(N-1)\frac{J^2}{16}+\frac{Nh_z^2}{4}\right]$. Finally, we also find
\begin{equation}
    \mathrm{Tr}[(AB)^2]=\frac{2^Nh_x^2}{4}\left[\frac{J^2}{16}(N - 1)(N - 4)+\frac{h_z^2}{4}N(N-2)\right]
\end{equation}

Next, we consider the XXZ model without NNN interaction, we have 
\begin{equation}
    A=\frac{J_1}{4}\sum_{j:\rm even}^{N-1}\left[\sigma_j^x\sigma_{j+1}^x+\sigma_j^y\sigma_{j+1}^y+\Delta_1\sigma_j^z\sigma_{j+1}^z\right]
\end{equation}
and
\begin{equation}
    B=\frac{J_1}{4}\sum_{j:\rm odd}^{N-1}\left[\sigma_j^x\sigma_{j+1}^x+\sigma_j^y\sigma_{j+1}^y+\Delta_1\sigma_j^z\sigma_{j+1}^z\right]
\end{equation}
We find the following traces
\begin{equation}
    \mathrm{Tr}[A^2]=\frac{J_1^22^N(2+\Delta_1^2)}{16}\sum_{j:\rm even}^{N-1}=\frac{J_1^22^N(2+\Delta_1^2)}{16}\left\lfloor\frac{N-1}{2}\right\rfloor
\end{equation}

\begin{equation}
    \mathrm{Tr}[B^2]=\frac{J_1^22^N(2+\Delta_1^2)}{16}\sum_{j:\rm odd}^{N-1}=\frac{J_1^22^N(2+\Delta_1^2)}{16}\left\lceil\frac{N-1}{2}\right\rceil
\end{equation}

\begin{equation}
    \mathrm{Tr}[A^2B^2]=\left(\frac{J_1}{4}\right)^4(2+\Delta_1^2)^22^N\left\lfloor\frac{N-1}{2}\right\rfloor\left\lceil\frac{N-1}{2}\right\rceil
\end{equation}

\begin{equation}
    \mathrm{Tr}[(AB)^2]=\left(\frac{J_1}{4}\right)^4(2+\Delta_1^2)^22^N\left\lfloor\frac{N-1}{2}\right\rfloor\left\lceil\frac{N-1}{2}\right\rceil-4\left(\frac{J_1}{4}\right)^4(1+2\Delta_1^2)2^N(N-2)
\end{equation}

and $\mathrm{Tr}[AB]=0$. Finally, we would like to include the NNN interaction and compute the relevant traces based on the three blocks $A$, $B$, and $C$ defined in Eqs. \eqref{eqn:XXZ_A}-\eqref{eqn:XXZ_C}. We list below all the traces appearing in our approximate analytical forms of the variational parameters
\begin{equation}
\label{eqn:firstXXZ_NNN_trace}
    \mathrm{Tr}[A^2]=\mathrm{Tr}[B^2]=\frac{2^N}{16}\left[J_1^2(N-1)+J_2^2(N-2)\right]
\end{equation}
\begin{equation}
    \mathrm{Tr}[C^2]=\frac{2^N}{16}\left[J_1^2\Delta_1^2(N-1)+J_2\Delta_2^2(N-2)\right]
\end{equation}
\begin{equation}
     \mathrm{Tr}[AB]=\mathrm{Tr}[AC]=\mathrm{Tr}[BC]=0
\end{equation}

\begin{equation}
    \mathrm{Tr}[A^2B^2]=2^{N-8}\left[J_1^2(N-1)+J_2^2(N-2)\right]^2
\end{equation}

\begin{equation}
    \mathrm{Tr}[A^2C^2]=\mathrm{Tr}[B^2C^2]=2^{N-8}
\left[
\Delta_1^{2}\!J_1^2(N-1)
+
\Delta_2^{2}\!J_2^2(N-2)
\right]
\left[
J_1^2(N-1)
+
J_2^2(N-2)
\right]
\end{equation}

\begin{equation}
    \mathrm{Tr}[A^2BC]=\mathrm{Tr}[A^2CB]=-\left(\frac{J_1}{4}\right)^2\frac{J_2}{4}\left[\frac{\Delta_2J_2}{2}+\Delta_1J_1\right]2^N(N-2)
\end{equation}

\begin{equation}
    \mathrm{Tr}[(AB)^2]=2^{N-8}\left[J_1^4(N-3)^2+2J_1^2J_2^2(N^2-11N+22)+J_2^4\begin{cases}
1, & \text{if } N=3, \\
N^2-8N+20, & \text{if } N>3.
\end{cases}\right]
\end{equation}

\begin{equation}
    \mathrm{Tr}[(AC)^2]=2^{N-8}\left[\Delta_1^2J_1^4(N-3)^2+(\Delta_1^2+\Delta_2^2)J_1^2J_2^2(N^2-11N+22)+\Delta_2^2J_2^4\begin{cases}
1, & \text{if } N=3, \\
N^2-8N+20, & \text{if } N>3.
\end{cases}\right]
\end{equation}

\begin{equation}
    \mathrm{Tr}[(BC)^2]=\mathrm{Tr}[(AC)^2]
\end{equation}
and
\begin{equation}
\label{eqn:lastXXZ_NNN_trace}
    \mathrm{Tr}[ABAC]=2^{N-7}(N-2)J_1^2J_2\left[2\Delta_1J_1+\Delta_2J_2\right]
\end{equation}

\subsection{Dynamical traces related to ansatzes Eqs. (\eqref{eqn:ansatz_V2}-\eqref{eqn:ansatz_V3}) for the quantum Ising model}
\label{appIsingTraceDynamical}
We start with the unitary ansatz in Eq. \eqref{eqn:ansatz_V2}. As shown in Appendix \ref{app:EOM}, a few time-dependent traces show up in the corresponding EOMs. The first was
\begin{equation}
    T_1(c_1)=\mathrm{Tr}\left[Ae^{ic_1B}Ae^{-ic_1B}\right],
\end{equation}
Below, we will show how this trace can be computed exactly regardless of the parameters or the number of qubits. 
For simplifications, we will work with Pauli matrices in which we write $A=\frac{h_x}{2}\sum_{j=1}^N\sigma_j^x$ and $B=\frac{J}{4}\sum_{j=1}\sigma_j^z\sigma_{j+1}^z+\frac{h_z}{2}\sum_{j=1}^N\sigma_j^z$. Next, we will use the $\sigma^z$ basis per qubit
\begin{equation}
    \sigma^z\ket{\pm}=\pm \ket{\pm}
\end{equation}
which forms a basis of size $2^N$ for the N qubit system. We label each basis state as $\ket{k}$. Thus, we rewrite $g_{02}$ as
\begin{equation}
    T_1(c_1)=\sum_k\langle k|Ae^{ic_1B}Ae^{-ic_1B}|k\rangle 
\end{equation}
Noting that
\begin{equation}
    e^{ic_1B}\ket{k}=e^{ic_1b_k}
\end{equation}
where $b_k=\frac{J}{4}\sum_{j=1}s_j^k s_{j+1}^k+\frac{h_z}{2}\sum_{j=1}^Ns_j^k$ with $s_j^k=\pm 1$. Thus, we have
\begin{equation}
    T_1(c_1)=\frac{h_x^2}{4}\sum_{k=1}^{2^N} \sum_{j,\ell=1}^N e^{-ic_1b_k} \langle k|\sigma_j^x e^{ic_1B}\sigma_\ell^x|k\rangle
\end{equation}
Noting that $\sigma_j^x$ flips the spin at site $j$, we define $\ket{k_j}$ to be the state obtained by flipping the spin at site $j$ in the state $\ket{k}$. Thus, we have
\begin{equation}
    T_1(c_1)=\frac{h_x^2}{4}\sum_{k=1}^{2^N} \sum_{j,\ell=1}^N e^{-ic_1b_k} \langle k_j| e^{ic_1B}|k_\ell\rangle
\end{equation}
Using the fact that $\ket{k_\ell}$ still an eigenstate of $B$ and that $\bra{k_j}k_\ell\rangle=\delta_{j\ell}$, we get
\begin{equation}
    T_1(c_1)=\frac{h_x^2}{4}\sum_{k=1}^{2^N} \sum_{j=1}^Ne^{ic_1(b_{k_j}-b_k)}=\frac{h_x^2}{4}\sum_{j=1}^N\sum_{k=1}^{2^N} e^{ic_1(b_{k_j}-b_k)}=\frac{h_x^2}{4}\sum_{j=1}^N Z_j
\end{equation}
where
\begin{equation}
    Z_j=\sum_{k=1}^{2^N} e^{ic_1(b_{k_j}-b_k)}
\end{equation}
Recalling that $b_k=\frac{J}{4}\sum_{\ell=1}s_\ell^k s_{\ell+1}^k+\frac{h_z}{2}\sum_{\ell=1}^Ns_\ell^k$, we find $b_{k_j}$ by simply flipping the sign of $s_j^k$, giving:
\begin{equation}
   b_{k_j}-b_k=\begin{cases}
       -\frac{J}{2}s_j^ks_{j+1}^k-h_zs_j^k & \text{if $j=1,N$}\\
       -\frac{J}{2}\left[s_{j-1}^ks_j^k+s_j^ks_{j+1}^k\right]-h_zs_j^k & \text{if $1<j<N$}.
   \end{cases} 
\end{equation}
For $j=1,N$, we find
\begin{align}
    Z_1=Z_N&=2^{N-2}\sum_{s_1,s_2=\pm 1}e^{-ic_1\left[\frac{J}{2}s_1s_{2}+h_zs_1\right]}=2^{N-1}\sum_{s_2=\pm 1}\cos\left[c_1\left(\frac{J}{2}s_{2}+h_z\right)\right]\nonumber\\
    &=2^{N-1}\left[\cos\left[c_1\left(\frac{J}{2}+h_z\right)\right]+\cos\left[c_1\left(\frac{J}{2}-h_z\right)\right]\right]
\end{align}
Now, for the middle sites $1<j<N$, we have
\begin{align}
    Z_j=\chi&=2^{N-3}\sum_{s_1,s_2,s_3=\pm 1}e^{-ic_1s_2\left[\frac{J}{2}\left(s_1+s_3\right)+h_z \right]}=2^{N-2}\sum_{s_1,s_3=\pm 1}\cos\left[c_1\left(\frac{J}{2}\left(s_1+s_3\right)+h_z \right)\right]\nonumber\\
    &=2^{N-2}\Big[\cos\left[c_1\left(J+h_z \right)\right]+2\cos\left[c_1h_z \right]+\cos\left[c_1\left(J-h_z \right)\right]\Big]
\end{align}
Thus, we have our final analytic expression
\begin{align}
    T_1(c_1)&=\frac{h_x^2}{4}(N-2)2^{N-2}\Big[\cos\left[c_1\left(J+h_z \right)\right]+2\cos\left[c_1h_z \right]+\cos\left[c_1\left(J-h_z \right)\right]\Big]\nonumber\\
    &+\frac{h_x^2}{4}2^{N}\left[\cos\left[c_1\left(\frac{J}{2}+h_z\right)\right]+\cos\left[c_1\left(\frac{J}{2}-h_z\right)\right]\right]\nonumber\\
    &=h_x^2 \, 2^{N-2} \cos(c_1 h_z) \cos\!\left(\frac{c_1 J}{2}\right) \left[ (N-2) \cos\!\left(\frac{c_1 J}{2}\right) + 2 \right]
\end{align}  
Our second dynamical trace is
\begin{equation}
    T_2(c_0)=\mathrm{Tr} \left[Be^{ic_0A}Be^{-ic_0A}\right]
\end{equation}
To make the analytic work smooth, we use the $\sigma^x$ basis
\begin{equation}
    \sigma^x\ket{x_{\pm}}=\pm\ket{x_\pm}
\end{equation}
where
\begin{equation}
    \ket{x_{\pm}}=\frac{\ket{+}\pm\ket{-}}{\sqrt{2}}
\end{equation}
In this case $\sigma^z\ket{x_{\pm}}=\ket{x_{\mp}}$ acting as a flipping operator. Using this, we can compute $T_2(c_0)$ as follows
\begin{align}
    T_2(c_0)&=\sum_{k=1}^{2^N}\bra{k}Be^{ic_0A}Be^{-ic_0A}\ket{k}=\sum_{k=1}^{2^N}e^{-ic_0a_k}\bra{k}Be^{ic_0A}B\ket{k}\nonumber\\
    &=\frac{J^2}{16}\sum_{k=1}^{2^N}\sum_{j,\ell=1}^{N-1}e^{-ic_0a_k}\bra{k_{j,j+1}}e^{ic_0A}\ket{k_{\ell,\ell+1}}+\frac{h_z^2}{4}\sum_{k=1}^{2^N}\sum_{j,\ell=1}^{N}e^{-ic_0a_k}\bra{k_j}e^{ic_0A}\ket{k_\ell}\nonumber\\
    &=\frac{J^2}{16}\sum_{k=1}^{2^N}\sum_{j=1}^{N-1}e^{ic_0(a_{k_{j,j+1}}-a_k)}+\frac{h_z^2}{4}\sum_{k=1}^{2^N}\sum_{j=1}^{N}e^{ic_0(a_{k_j}-a_k)}
\end{align}
where $a_k=\frac{h_x}{2}\sum_{p=1}^Ns_p$ where $s_p=\pm1$, $a_{k_j}$ obtained from $a_k$ by flipping the spin at site $j$, and $a_{k_{j,j+1}}$ is found from $a_k$ by flipping the spin at sites $j$ and $j+1$ (in the $\sigma^x$ basis). Noting that $a_{k_{j,j+1}}-a_k=-h_x(s_p+s_{p+1})$, we write the first term in $T_2(c_0)$ as
\begin{align}
    \frac{J^2}{16}\sum_{k=1}^{2^N}\sum_{j=1}^{N-1}e^{ic_0(a_{k_{j,j+1}}-a_k)}&=\frac{J^2}{16}\sum_{j=1}^{N-1}\sum_{k=1}^{2^N}e^{ic_0(a_{k_{j,j+1}}-a_k)}=\frac{J^22^{N-2}}{16}\sum_{j=1}^{N-1}\sum_{s_p,s_{p+1}=\pm 1}e^{-ic_0h_x(s_p+s_{p+1})}\nonumber\\
    &=\frac{J^22^{N}(N-1)}{16}\cos^2(h_xc_0)
\end{align}
Similarly, the second term becomes
\begin{equation}
    \frac{h_z^2}{4}\sum_{k=1}^{2^N}\sum_{j=1}^{N}e^{ic_0(a_{k_j}-a_k)}=\frac{h_z^2}{4}\sum_{j=1}^{N}\sum_{k=1}^{2^N}e^{ic_0(a_{k_j}-a_k)}=\frac{h_z^22^{N-1}}{4}\sum_{j=1}^{N}\sum_{s_p=\pm1}e^{-ic_0h_x s_p}=\frac{h_z^22^{N}N}{4}\cos(h_x c_0) 
\end{equation}
Thus, we have
\begin{equation}
    T_2(c_0)=\frac{J^22^{N}(N-1)}{16}\cos^2(h_xc_0)+\frac{h_z^22^{N}N}{4}\cos(h_x c_0) 
\end{equation}
The last trace that appears in this ansatz is
\begin{equation}
    T_3(c_0,c_1)=\mathrm{Tr} \left[e^{ic_1B}Ae^{-ic_1B}e^{-ic_0A}Be^{ic_0A}\right]
\end{equation}
Following the same steps above, we find
\begin{align}
T_3(c_0,c_1)&=h_x 2^{N-2} \sin(h_x c_0) \Biggl\{ 
h_z \sin(c_1 h_z) \cos\!\left(\frac{c_1 J}{2}\right) \left[2 + (N-2)\cos\!\left(\frac{c_1 J}{2}\right)\right]\nonumber \\
&\qquad + J \cos(h_x c_0) \cos(c_1 h_z) \sin\!\left(\frac{c_1 J}{2}\right) \left[1 + (N-2)\cos\!\left(\frac{c_1 J}{2}\right)\right] \Biggr\},
\end{align}
Thus, with those traces we have obtained a complete analytic expressions of $g$ and $F$ for the quantum Ising model. These expressions valid in the thermodynamic limit and at arbitrary coupling parameters substantially reducing the computational overhead of matrix exponentials which is not also expensive but becomes infeasible at large $N$. 

Secondly, using the ansatz in Eq. \eqref{eqn:ansatz_V3}, two new dynamical trace emerges. For the ABAB operator ordering we encounter the following trace
\begin{equation}
T_4(c_0,c_1,c_2)=\mathrm{Tr}\left[e^{ic_2A} B e^{-ic_2A} e^{-ic_1B} e^{-ic_0A} B e^{ic_0A} e^{ic_1B}  \right]
\end{equation}
After algebraic steps, we obtain the following analytic form of $T_4(c_0,c_1,c_2)$ 
\begin{equation}
\begin{aligned}
\frac{T_4(c_0,c_1,c_2)}{2^N}&=\frac{Nh_z^2\cos(2\theta)\cos(2\phi)}{4}+\frac{(N-1)J^2\cos^2(2\theta)\cos^2(2\phi)}{16}\\
&+\frac{Jh_z\sin(2\theta)\sin(4\phi)\sin(2\gamma)\sin(2\lambda)}{8}\Big[1+(N-2)\cos(2\lambda)\Big]\\
&-\frac{h_z^2\sin(2\theta)\sin(2\phi)\cos(2\gamma)\cos(2\lambda)}{4}\Big[2+(N-2)\cos(2\lambda)\Big]\\
&+\frac{J^2\sin^2(2\theta)\sin^2(2\phi)\cos^2(2\gamma)\cos(2\lambda)}{16}\Big[2+(N-3)\cos(2\lambda)\Bigg]\\
&+\frac{Jh_z\sin(4\theta)\sin(2\phi)\sin(2\gamma)\sin(2\lambda)}{8}\Big[1+(N-2)\cos(2\lambda)\Big]\\
&-\frac{J^2\sin(4\theta)\sin(4\phi)\cos(2\gamma)}{32}\Big[\cos(2\lambda)+(N-2)\cos(4\lambda)\Big],
\end{aligned}
\end{equation}
where
\begin{equation}
    \theta=\frac{h_xc_2}{2},\quad\phi=\frac{h_xc_0}{2},\quad\gamma=\frac{h_zc_1}{2},\quad\lambda=\frac{Jc_1}{4}
\end{equation}

while for the other operator ordering (BABA) we come across the following trace:
\begin{equation}
T_5(c_0,c_1,c_2)=\mathrm{Tr}\left[e^{ic_2B} A e^{-ic_2B} e^{-ic_1A} e^{-ic_0B} A e^{ic_0B} e^{ic_1A}  \right]
\end{equation}
which we find its analytic formula to be
\begin{equation} 
\begin{aligned}
\frac{T_5(c_0,c_1,c_2)}{2^N}&=\frac{h_x^2\cos(2\theta')\cos(2\phi')}{4}\Big[2\cos(2\lambda')\cos(2\gamma')+(N-2)\cos^2(2\gamma')cos^2(2\lambda')\\
&-2\sin(2\lambda')\left(\sin(2\gamma')\cos^2(2\omega')-\frac{\sin(4\gamma')\sin^2(2\omega')}{2}\right)\\
&+\sin(4\lambda')\left(\sin(2\gamma')\sin^2(2\omega')-\frac{\sin(4\gamma')\cos^2(2\omega')}{2}\right)\\
&-\frac{(N-3)\sin(4\lambda')\sin(4\gamma')\cos(4\omega')}{2}+(N-2)\sin^2(2\lambda')\cos^2(2\omega')\sin^2(2\gamma')\Big]\\
&-\frac{h_x^2\sin(2\theta')\sin(2\phi')}{4}\Big[2\cos(2\lambda')\cos(2\gamma')\cos(2\omega')+(N-2)\cos^2(2\lambda')\cos^2(2\gamma')\cos(2\omega')\\
&-2\sin(2\lambda')\sin(2\gamma')\cos(2\omega')-\frac{(N-2)\sin(4\lambda')\sin(4\gamma')\cos(2\omega')}{2}\\
&+(N-2)\sin^2(2\lambda')\sin^2(2\gamma')\cos^3(2\omega')\Big],
\end{aligned}
\end{equation}
where
\begin{equation}
    \theta'=\frac{h_zc_2}{2},\quad \lambda'=\frac{Jc_2}{4},\quad \phi'=\frac{h_zc_0}{2},\quad \gamma'=\frac{Jc_0}{4}, \quad \omega=\frac{h_xc_1}{2}.
\end{equation}  

\end{widetext}

\bibliography{ref.bib}
\end{document}